\documentclass[twocolumn,aps,preprintnumbers,epsfig,nofootinbib]{revtex4}

\usepackage{graphicx}
\usepackage{epstopdf}
\usepackage{latexsym}
\usepackage{amssymb}
\usepackage{amsmath}
\usepackage{color}
\usepackage{mathrsfs}

\usepackage[center]{subfigure}

\begin{document}

 \preprint{YITP-17-36,  IPMU17-0056}

\title{No static black hole hairs in gravitational  theories with broken Lorentz invariance}

\author{Kai Lin $^{a, b}$}

\author{Shinji Mukohyama $^{c, d}$}

\author{Anzhong Wang $^{a, e}$}

\author{Tao Zhu $^{a, e}$}

\affiliation{$^{a}$  Institute  for Advanced Physics $\&$ Mathematics, Zhejiang University of Technology, Hangzhou 310032,  China\\
$^{b}$ Instituto de F\'isica e Qu\'imica, Universidade Federal de Itajub\'a, CEP 37500-903,  Itajub\'a, MG, Brazil\\
$^{c}$ Center for Gravitational Physics, Yukawa Institute for Theoretical Physics, Kyoto University, 606-8502, Kyoto, Japan\\
$^{d}$ Kavli Institute for the Physics and Mathematics of the Universe (WPI), The University of Tokyo Institutes for Advanced Study, 
The University of Tokyo, Kashiwa, Chiba 277-8583, Japan\\
 $^{e}$ GCAP-CASPER, Physics Department, Baylor University, Waco, TX 76798-7316, USA
 }

\date{\today}

\begin{abstract}

In this paper, we revisit the issue of static hairs of black holes in gravitational theories with broken Lorentz invariance in the case that the speed 
$c_{\phi}$ of the khronon field becomes infinitely large, $c_{\phi} = \infty$, for which the sound horizon of the khronon field coincides with the 
universal horizon, and the boundary conditions at the sound horizon reduce to those given normally at the universal horizons. As a result, less 
boundary conditions are present in this extreme case in comparison with the case $c_{\phi} = $ finite. Then, it would be expected that static hairs 
might exist. However, we show analytically that even in this case static hairs still cannot exist, based on a decoupling limit analysis. We also 
consider the cases in which $c_{\phi}$ is finite  but with  $c_{\phi} \gg 1$, and obtain the same conclusion.

\end{abstract}


\maketitle

\section{ Introduction }
\renewcommand{\theequation}{1.\arabic{equation}} \setcounter{equation}{0}

The studies of black holes in gravitational theories with broken Lorentz invariance (LI)  have attracted lots of
attention recently \cite{UHs}, as it was generally believed
that black holes  in such theories might be  only the low-energy phenomena and  do not exist at all in the ultraviolet (UV).
This is mainly because, once Lorentz symmetry is broken,  particles from different species can have different
speeds, and in principle the speeds can be arbitrarily large. Then, intuition tells us that particles with  sufficient  high speeds
can always escape to infinity no matter how they are close to  singularities.

In the infrared (IR), LI must be  restored   in the matter sector. Otherwise, it will be in serious conflict with
observations \cite{LSB}. In fact, no matter how high the scale of the LI breaking is,
it always has  significant effects on the low-energy physics   \cite{Collin04}.
However,  in the gravitational sector, observational constraints of Lorentz violations are rather weak  \cite{LSB}, and
there exist various (low-energy)   theories in which the Lorentz symmetry is spontaneously broken  by the development of non-zero vacuum expectation values,
for example, the ghost condensation \cite{GC},  in which the LI is broken by the development of the non-zero vacuum expectation values of the derivative of a scalar field. 
 In the Einstein-aether theory \cite{EA},  it is broken by the existence of a globally timelike aether field. Because of such a breaking, extra gravitational modes exist.
In particular, in the Einstein-aether theory in addition to the spin-2 mode, spin-0 and spin-1 modes also appear, which all move in different speeds \cite{EA}.
Then, a black hole horizon would appear larger for subluminal particles and smaller for superluminal ones. As a result, the temperature of the Hawking radiation will be different 
for different species of particles. Dubovsky and Sibiryakov (DS) showed  that in such cases  one can always construct thought experiments, such as the perpetuum mobile of the second kind,
in which the back hole mass and angular momentum remains constant, while the entropy outside the black hole decreases, whereby the generalized second law of BH thermodynamics  is violated 
\cite{DS06}. This poses a great challenge to any theories with broken LI.
Later, Eling et al generalized DS' results to more general cases by considering both of the possibilities of the Penrose process and semiclassical heat flow
 \cite{EFJW}. However, in the ghost condensation, it was  shown that the second law can be saved after the accretion of ghost condensation
 is taken into account  \cite{Mukohyama:2009rk,Muk10}. (For consideration of de Sitter thermodynamics in the context of ghost inflation, see \cite{Jazayeri:2016jav}.)

Recently, a potential breakthrough  was the discovery  \cite{BS11,BJS} that   there still exist absolute causal boundaries, the so-called {\em universal horizons},  in gravitational
theories  with broken LI.  Particles even with infinitely large velocities    would just move around on these boundaries and  cannot escape to infinity.  The discovery is born out of
the realization that causal structures of spacetimes in such theories are quite different from those given in general relativity (GR).
For example, in Ho\v{r}ava gravity \cite{Horava}, the theory is gauge-invariant under  the foliation-preserving diffeomorphism,
 \begin{equation}
\label{0.1}
 t' =  f(t), \quad {x'}^i =  \xi^i\left(t, x^k\right),\; (i = 1, 2, 3)
\end{equation}
which clearly breaks the (local and global) LI, 
\begin{equation}
\label{0.1a}
{x'}^{\mu} = L^{\mu}_{\nu}x^{\nu}, \;  (\mu, \nu = 0, 1, 2, 3). 
\end{equation}
As a result,   the dispersion relation becomes generically nonlinear
(See, for example,  \cite{WM10} and references therein),
 \begin{eqnarray}
 \label{0.2}
E^2&=& c_p^2p^2\left[1+\alpha_1\left(\frac{p}{M_*}\right)^2+\alpha_2\left(\frac{p}{M_*}\right)^4 \right.\nonumber\\
&& ~~~~~~~~ \left. + ... +\alpha_{d-1}\left(\frac{p}{M_*}\right)^{2(d-1)}\right],
 \end{eqnarray}
where, $\alpha_i$ and $c_p$ are coefficients, depending on the particular specie of the particle, and $E$ and $p$ are the energy
and momentum of the particle considered, while $M_*$ is the suppression energy scale of the higher-order operators in the
theory. Then, both of the group and phase velocities become unbounded as $p$ becomes larger and larger.  Due to the presence of particles with
arbitrarily large speeds,  the causal structure of a spacetime  is quite different from that of GR, but very similar to that of Newtonian theory    \cite{GLLSW}. 
In particular, for a given set of events $p$ and $q$, the notion of past and future is uniquely determined by the time difference, $\Delta{t} \equiv t_{p} - t_{q}$, 
between the two events.   If $\Delta{t} > 0$, the event $q$ is to the past of $p$; if $\Delta{t} < 0$, it  is to the future; and if $\Delta{t} = 0$, the two events are 
simultaneous.  Similar to the Newtonian case, the causality is achieved by assuming that particles are always moving along the increasing direction of $t$.

\begin{figure}[tbp]
\centering
\includegraphics[width=1\columnwidth]{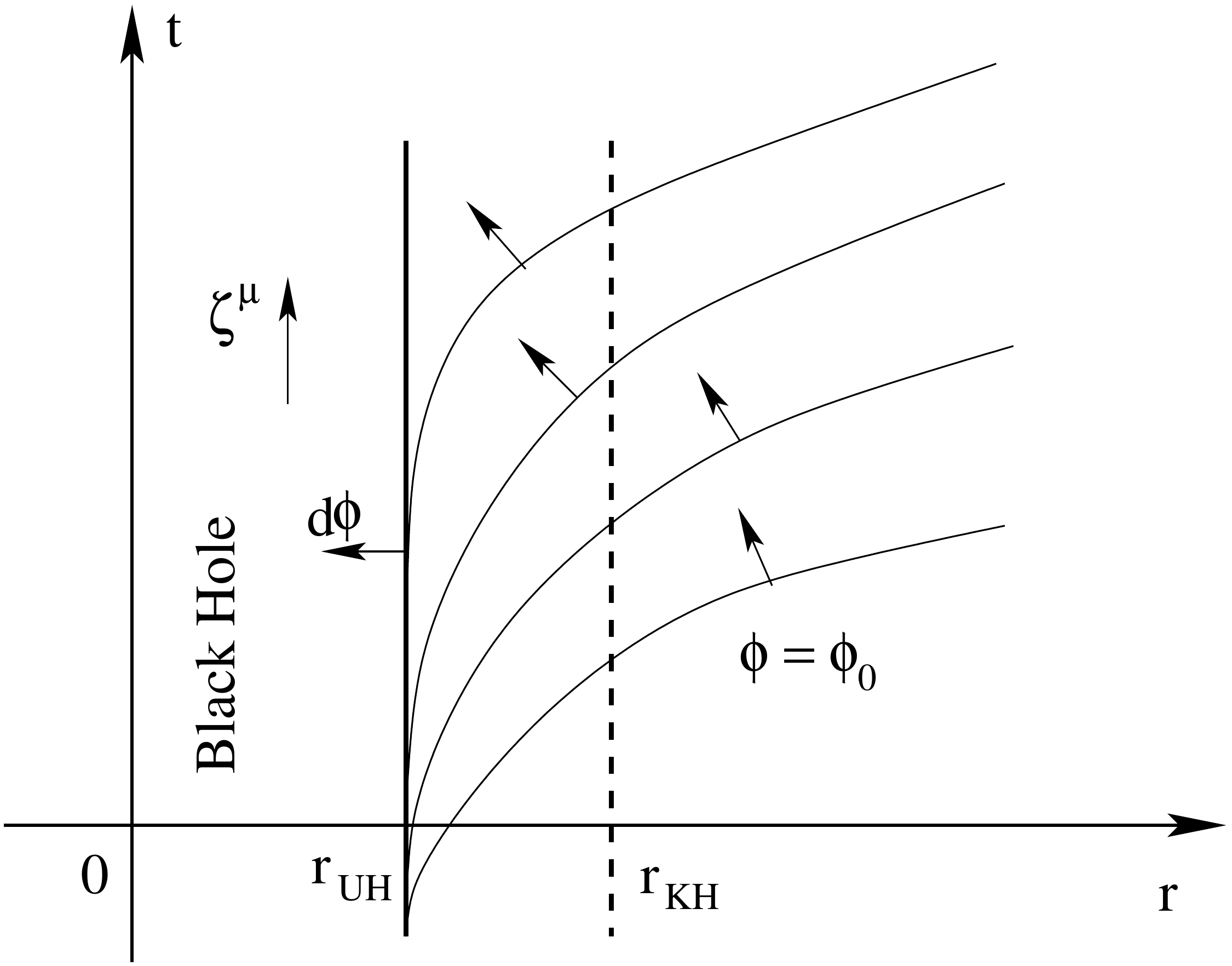}
\caption{Illustration of the bending of the $\phi$ = constant surfaces, and the existence of the universal horizon  in the Schwarzschild spacetime,
where $\phi$ denotes the  globally timelike scalar  field, and $t$ and $r$ are the Painlev\'{e}-Gullstrand coordinates. Particles move always along the increasing direction of $\phi$.
The Killing vector $\zeta^{\mu} = \delta^{\mu}_{t}$ always points upward at each point of the plane. The vertical dashed  line is   the location of the Killing horizon,
$r=r_{KH}$. The universal horizon, denoted by the vertical solid  line, is  located at $r = r_{UH}$, which is always inside the Killing horizon \cite{LSW16}. }
\label{Fig0}
\end{figure}

The diffeomorphism invariance, like any gauge symmetry, is a redundancy of descriptions and thus can be restored by introducing extra degrees of freedom. 
The khronometric  theory  \cite{BPS} is one of such examples and thus is gauge-invariant under the full general diffeomorphism,
 \begin{equation}
\label{0.3}
 \delta{x'}^{\mu} =  \xi^{\mu}\left(t, x^k\right).
\end{equation}
The  khronon field $\phi$ always develops a non-zero vacuum expectation value of derivative that is always time-like,
\begin{equation}
\label{0.4}
u_{\mu} \equiv  \frac{\partial_{\mu} \phi}{\sqrt{X}}, \quad X \equiv -g^{\alpha\beta}\partial_{\alpha} \phi \partial_{\beta} \phi > 0,
\end{equation}
whereby a special time-like direction is created.  Assume that all particles  move along the increasing direction of $\phi$, so that the causality of a given spacetime is
still well defined, similar to the Newtonian case. In such a spacetime, there may exist a surface as shown in  Fig.~\ref{Fig0}, denoted by the vertical solid line,
located at $r = r_{UH}$. Given that all particles move along the increasing direction of the time-like scalar field, from Fig.~\ref{Fig0} it is clear that a particle must cross this surface and move inward,
once it arrives at it, no matter how large of its speed is. This is a one-way membrane, and particles even with infinitely large speed cannot escape from it, once they are
inside it. So, it acts as a universal horizon to all particles (with any speed)  \cite{BS11,BJS}.  At the universal horizon, we have
$\left(dt\cdot d\phi\right) = 0$, or equivalently,
\begin{equation}
\label{0.5}
\left(\zeta \cdot u\right)   = 0,
\end{equation}
where $\zeta \equiv \partial_t$ denotes the asymptotically time-like Killing vector and $u \equiv u_{\lambda}dx^{\lambda}$ is the four-velocity of the khronon field.

It should be noted that the concept of universal horizons was first proposed in the studies of black holes in the Einstein-aether theory \cite{EA}, which is essentially a vector-tensor theory of gravity but with
the vector being always time-like. When the vector is hypersurface-orthogonal, that is, when it satisfies the conditions,
\begin{equation}
\label{0.6}
u_{[\alpha}D_{\beta}u_{\lambda]} = 0,
\end{equation}
it can be shown that there at least locally exists a scalar field $\phi$ whose derivative is time-like so that $u_{\mu}$ is given by Eq.(\ref{0.4}) \cite{Wald94}. Then, the khrononmetric theory is equivalent to 
the  hypersurface-orthogonal Einstein-aether theory in the level of action, as shown explicitly by Jacobson in  \cite{Jacob}. It must be noted that fundamentally  these are two different theories \cite{Wang12}. 
In particular, in the khrononmetric theory an instantaneous mode exists, which is absent in the Einstein-aether theory. It is the existence of this mode that might resolve the problems mentioned above \cite{BS11}. 
For more details, we refer readers to \cite{Wang17}.

It is also exactly because of this that motivates us to revisit the problem of the existence of static hair in the extreme case $c_{\phi} = \infty$.

By definition the khronon field is part of the theories. In particular, it represents  an extra degree of freedom in the gravitational sector, in addition to the spin-2 graviton appearing in GR. 
In generic situations one thus needs to analyze the whole set of gravitational field equations to investigate the nature of universal horizons. This is a complicated task in general. However, 
there is a useful decoupling limit in which the analysis is significantly simplified~\cite{BS11}. The decoupling limit is achieved by supposing that the coupling constants $c_{1,\cdots,4}$ 
(see (\ref{1.12})) of the khronon field are small, and is motivated by observational constraints on the theories~\cite{BPS}. In the decoupling limit the stress-energy tensor of the khronon
 field is small and thus we can safely ignore the backreaction of the khronon field to the geometry. In other words, we can treat the khronon field as a test field on a fixed geometry that is 
 a solution to the gravitational field equations without the khronon field \cite{LACW}. In the decoupling limit it was shown that universal horizons exist in the three well-known  solutions:  
 the Schwarzschild, Schwarzschild anti-de Sitter, and  Reissner-Nordstr\"om \cite{LGSW}, which are also solutions of  Ho\v{r}ava gravity in a proper choice of coordinates  \cite{GLLSW,Wang17}.

A further generalization to   spacetimes with a preferred foliation (\ref{0.1}) was carried out systematically \cite{BCS16}, and shown that, among other things, the zeroth law of black hole mechanics holds at universal horizons.

Regarding the first law, the situation seems more complicated. It was first shown that  it holds  \cite{BBMa}, provided that   the surface gravity is  now defined as  \cite{CLMV}
 \begin{eqnarray}
 \label{0.7}
\kappa_{UH} \equiv  \frac{1}{2} u^{\alpha} D_{\alpha} \left(u_{\lambda} \zeta^{\lambda}\right),
 \end{eqnarray}
which was obtained by considering the peeling behavior of ray trajectories of constant khronon field $\phi$  \cite{CLMV}.
Furthermore,   the considerations of Hawking radiation at the universal horizon also supports such a conclusion \cite{BBMb}. However, this seems true only for neutral case. When the black hole is charged,
 such a law is still absent \cite{DWW}.

 In addition,   using the Hamilton-Jacobi  method, quantum tunneling of   non-relativistic particles with a general non-linear dispersion relation was studied \cite{DWWZ}, 
 and it was found  that  different species of particles in general experience different temperatures,
\begin{equation}
\label{0.8}
T_{UH}^{z\ge 2} = \frac{2(z-1)}{z}\left(\frac{\kappa_{UH}}{2\pi}\right),
\end{equation}
where $\kappa_{UH}$ is the surface gravity calculated from Eq.(\ref{0.7}) and $z$ is the exponent of the dominant term in the UV.  When  $z = 2$ it reduces to the case considered in  \cite{BBMa,CLMV}.
Recently,  more careful studies of ray trajectories showed that the surface gravity for particles with a general non-relativistic dispersion relation  is given by \cite{Cropp16,DL16},
\begin{equation}
\label{0.9}
\kappa_{UH}^{z\ge 2} = \left(\frac{2(z-1)}{z}\right) \kappa_{UH}.
\end{equation}
 
In this paper, we would like to revisit the issue of the existence/absence of long static hair of black holes in the extreme case $c_{\phi} = \infty$. To recover the second law,
Blas and Sibiryakov  (BS) considered two possibilities where the missing entropy can be found  \cite{BS11}:
(i) It is accumulated somewhere inside the black hole (BH). BS studied the stabilities of the universal horizons and found that they are linearly stable.
 But, they argued that after nonlinear effects are taken into account, these horizons will be turned into singularities with
 finite areas. One hopes that in the full theory of Ho\v{r}ava  gravity this singularity is resolved into a high-curvature region of finite
width accessible to the instantaneous and fast high-energy modes. In this way the BH thermodynamics can be saved.
(ii) A BH has a large amount of static long hairs, which have tails that can be measured outside the horizon \cite{DTZ}. After measuring
them, an outer observer could decode the entropy that had fallen into the BH. In fact,  the processes described in \cite{DS06,EFJW} not only reduce the entropy outside of the horizon
but also change the state inside the black hole. Thus, the violation of the second law can be avoided, provided that this change is observable from outside by monitoring these hairs, in addition
to the mass and angular momentum of the black hole \cite{DTZ}. But, BS found that spherically symmetric hairs do not exist, and to have this
scenario to work, one has to use non-spherically symmetric hairs. 

However,  in the analysis of BS, the condition $c_{\phi} = $ finite was implicitly imposed \cite{BS11}. Then, the sound horizon of the khronon field will be different from the universal horizon, 
and the regularity conditions of the perturbations imposed at these horizons are also different and independent. Therefore, in the case $c_{\phi} = $ finite three independent sets of boundary
conditions must be imposed at the spatial infinity (in the case where the spacetime is asymptotically flat), the sound horizon, and the universal horizon,  respectively.  However, in the extreme 
case $c_{\phi} = \infty$, the sound horizon always coincides with the universal horizon, and the two sets of boundary conditions imposed at these horizons reduce to one. As a result, in the extreme 
case, we have less boundary conditions than those in the case $c_{\phi} = $ finite. So, an immediate question is: can static hairs exist in the extreme case? 

In this paper, we  shall investigate this issue, and show analytically that  long static hairs still do not exist even when  $c_{\phi} = \infty$.

The rest of the paper is organized as follows:  In Sec. II we give a brief review of universal horizons in static spacetimes, while in Sec. III we study static hairs, respectively,
 in  the Schwarzschild, Schwarzschild anti-de Sitter and Reissner-Nordstr\"om (RN) Backgrounds, and show analytically  that such hairs do not exist even when  $c_{\phi} = \infty$.
  It should be noted that in the RN case, our proof is restricted to the case $c_3 = 0$, and for the general case such a proof is still absent, 
  although it is not difficult to argue that it is more likely that the hairs do not exist even for the general case. 
  In Sec. IV, we study the large $c_{\phi}$ expansions, by assuming that $c_{\phi}$ is very large but finite. In this case, the boundary conditions
 across the sound and universal horizons, together with the ones at infinity, generically limit the existence of such hairs. The latter is consistent with the results
 obtained  in \cite{BS11}.  In Sec. V, we derive our main conclusions. In Appendix A, we present another derivation of analytical solutions of Eq.(\ref{II8}), and obtain the same null results, as
 expected.  In  Appendix B, we present some useful mathematical formulas for the study of the perturbations in the case where  $c_{\phi}$ is very  large but finite.

 \section{ Universal Horizons and Black Holes in Static Spacetimes}
\renewcommand{\theequation}{2.\arabic{equation}} \setcounter{equation}{0}

 In this section, we shall give a brief introduction of the universal horizons in static spacetimes. For detail, we refer readers to
 \cite{Wang17}.

 In terms of the Eddington-Finkelstein coordinates, static spacetimes are described by the metric,
 \begin{equation}
 \label{a.1}
 ds^2 = - F(r) dv^2 + 2f(r) dv dr + r^2d\Omega_k^2,
 \end{equation}
 where $ k = 0, \pm 1$, and
 \begin{eqnarray}
\label{2.2}
 d\Omega^2_k =\left\{
  \begin{array}{cc}
    d\theta^2+\sin^2\theta d\varpi^2,  &$k = 1$, \\
    d\theta^2+d\varpi^2,                & $k = 0$, \\
    d\theta^2+\sinh^2\theta d\varpi^2, & $k = -1$. \\
  \end{array}
\right.
 \end{eqnarray}
 In these coordinates, the time-translation Killing vector $\zeta^{\mu}$ is given by  $ \zeta^{\mu} = \delta^{\mu}_{v}$,
 and the location of the Killing horizons,  on which $\zeta^{\mu}$ becomes null, $ \left. \zeta^{\lambda} \zeta_{\lambda} \right|_{r = r_{EH}} = 0$,
 are the roots of the equation, $F(r) = 0$.

The existence of a universal horizon is closely related to the presence of the khronon field, which obeys the khronon equation \cite{Wang12,Wang17},
 \begin{eqnarray}
\label{1.11}
D_{\mu} {\cal{A}}^{\mu}  = 0,
\end{eqnarray}
where
\begin{eqnarray}
\label{1.12}
{\cal{A}}^{\mu} &\equiv& \frac{\left(\delta^{\mu}_{\nu}  + u^{\mu}u_{\nu}\right)}{\sqrt{X}}\AE^{\nu},\nonumber\\
\AE^{\nu} &\equiv& D_{\gamma} J^{\gamma\nu} + c_4 a_{\gamma} D^{\nu}u^{\gamma},\nonumber\\
J^{\alpha}_{\;\;\;\mu} &\equiv&  \big(c_1g^{\alpha\beta}g_{\mu\nu} + c_2 \delta^{\alpha}_{\mu}\delta^{\beta}_{\nu}
+  c_3 \delta^{\alpha}_{\nu}\delta^{\beta}_{\mu}\nonumber\\
&&  ~~~ - c_4 u^{\alpha}u^{\beta} g_{\mu\nu}\big)D_{\beta}u^{\nu}.
\end{eqnarray}
Here $a_{\mu} \equiv u^{\alpha}D_{\alpha}u_{\mu}$, $c_i$'s are coupling constants.
Among the three parameters, $c_1,\; c_3$ and $c_4$, only two of them
are independent.

Eq.(\ref{1.11}) is a second-order differential equation for $u_{\mu}$, and to uniquely determine it, two boundary conditions are needed.
These two conditions in stationary and asymptotically flat  spacetimes can be chosen as follows \cite{BS11} \footnote{These conditions can be easily generalized to
asymptotically  anti-de Sitter  spacetimes.}:

(i)  $u_{\mu}$ is
aligned asymptotically with the time translation Killing vector $\zeta_{\mu}$,
\begin{equation}
\label{1.13}
u^{\mu} \propto \zeta^{\mu}.
\end{equation}
(ii) The khronon has a regular future sound horizon, which
  is a null surface of the effective metric \cite{EJ},
\begin{equation}
\label{1.14}
g^{(\phi)}_{\mu\nu} = g_{\mu\nu} - \left(c_{\phi}^2 -1\right)u_{\mu}u_{\nu},
\end{equation}
where $c_{\phi}$ denotes the speed of the khronon  given by,
$c_{\phi}^2 = {c_{123}}/{c_{14}}$, where $c_{123}\equiv c_1+c_2+c_3,\; c_{14}\equiv c_1+c_4$.

A  universal horizon is defined by Eq.(\ref{0.5}).
Since $u_{\mu}$ is timelike globally, Eq.(\ref{0.5}) is possible only when $\zeta_{\mu}$ becomes spacelike. This can happen  only  inside Killing
 horizons. Then, we can define  the region inside the universal horizon as a black hole,
 since any signal   cannot escape to infinity, once it is trapped inside  it, no matter how large its velocity is.

In the case $f(r) = 1$, the khronon equation (\ref{1.11}) has a solution for $c_{14} = 0$, and given by \cite{LGSW} as
\begin{equation}
\label{2.10}
V =  -\frac{r_o^2}{r^2},\; (f = 1,\; c_{14} = 0),
\end{equation}
where $V=u^r$ and $r_o$ is an integration constant. Then, it can be shown  that universal horizons exist in the well-known black holes, the Schwarzschild, Schwarzschild anti-de Sitter, and Reissner-Nordstr\"om.

\section{ Linear  Perturbations of the Khronon Field and Static Hairs}
\renewcommand{\theequation}{3.\arabic{equation}} \setcounter{equation}{0}

\begin{figure}[tbp]
\centering
\includegraphics[width=8cm]{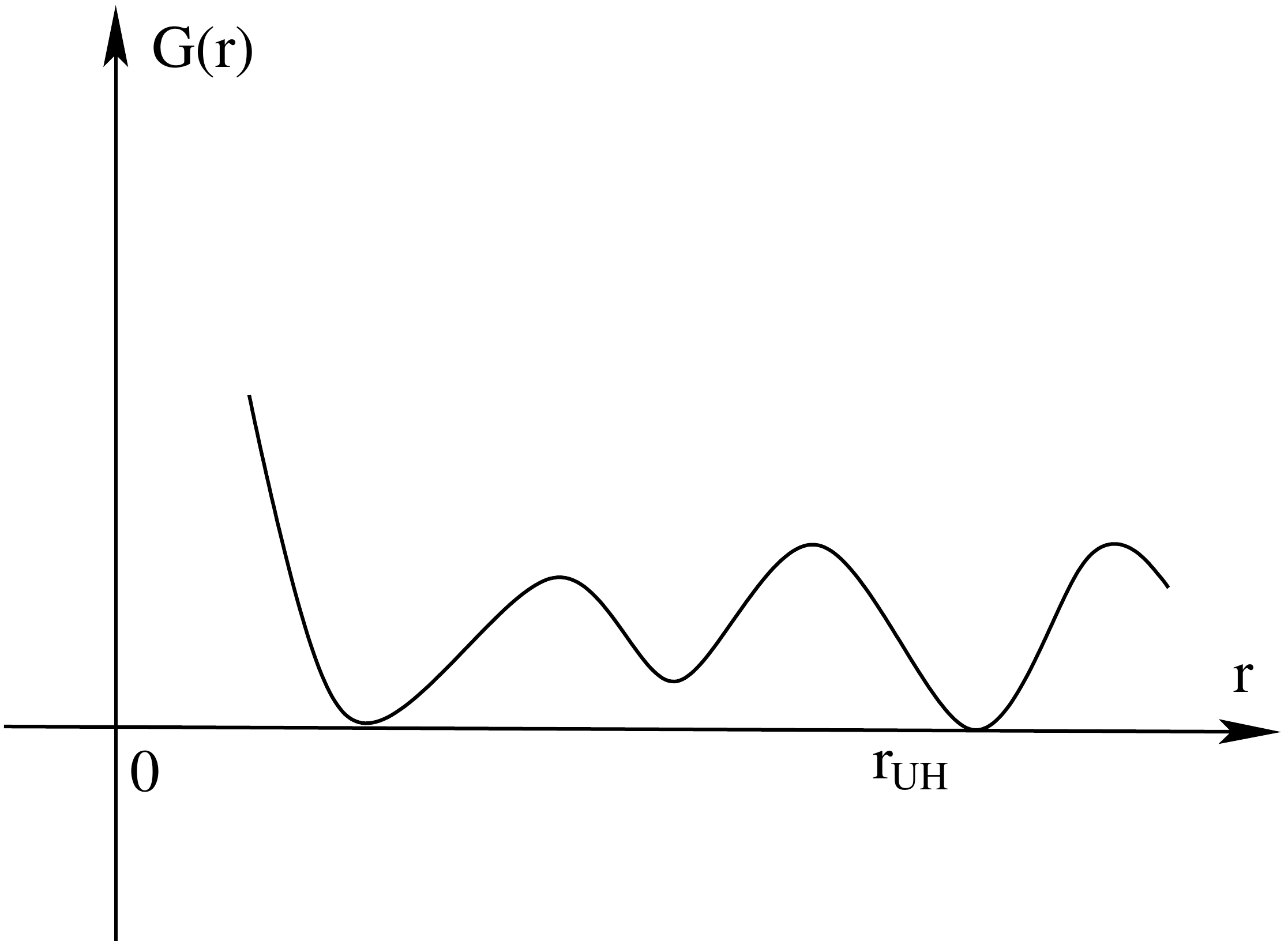}
\caption{ The general behavior of the functions $G(r) \equiv V^2(r) + F(r)$.} \label{fig1}
\end{figure}

At the linear level, if such hairs exist, they should manifest themselves in the form of regular static perturbations of the khronon \cite{DTZ}. In \cite{BS11},
 it was shown that they do not exist for finite $c_{\phi}$. So, in the rest of this paper we shall consider only the case $c_{\phi} = \infty$, i.e. $c_{14}=0$. As 
 mentioned before, this corresponds to the case where the sound horizon of the khronon coincides with the universal horizons. Since only two among ($c_1$, $c_3$, $c_4$) 
 are independent, in the following we shall choose $c_1 = c_4 = 0$ without loss of generality. From now on we also assume that 
\begin{equation}
\label{3.0}
f(r) = 1,\;\;\; k = 1.
\end{equation}
Then, we find that
\begin{equation}
\label{2.14}
\left(u\cdot\zeta\right) =    \sqrt{V^2 + F} \equiv U(r).
\end{equation}
 Thus, the universal horizon is located at   $U(r) = 0$. In general $U(r)$ can have several zero
points,  and we shall define the one with maximal radius as the universal horizon. In addition,  in order for the khronon field $\phi$ to be well-defined,   we must assume
\begin{eqnarray}
\label{2.13}
 G(r) \equiv V^2 + F\ge 0,
\end{eqnarray}
in the whole space-time, including the internal region of the Killing horizon, in which we have $F(r) < 0$. Therefore,   the universal horizon located at $
G(r) = 0$ must be also a minimum of $G(r)$, as illustrated in Fig.\ref{fig1}.
Thus, at the universal horizons $r= r_{UH}$ we have
\begin{equation}
\label{2.15}
\left. G(r) \right|_{r= r_{UH}} = 0 = \left. G'(r)\right|_{r= r_{UH}}.
\end{equation}

To proceed further, we introduce the coordinates $\tau$ and $\xi$ via the relations,
 \begin{equation}
 \label{III2}
\tau = v + \int{\frac{V+U}{\xi^2 UF}d\xi},~~~~\xi=\frac{1}{r}.
 \end{equation}
Then, in terms of $\tau$ and $\xi$, the metric (\ref{a.1}) with the conditions (\ref{3.0})  takes the form,
  \begin{eqnarray}
 \label{III3}
ds^2&=& -F(\xi)d\tau^2+2\frac{\sqrt{U(\xi)-F(\xi)}}{\xi^2U(\xi)}d\tau d\xi\nonumber\\
&& +\frac{d\xi^2}{\xi^4U(\xi)^2}+\frac{ d\Omega^2}{\xi^2}.
 \end{eqnarray}
 Clearly, this metric becomes singular at the universal horizon $U(\xi) = 0$, but regular across the Killing horizon $F(\xi) = 0$. To our current purpose, this is enough,
 as we shall impose the boundary conditions of the perturbations at both of the universal horizon and the spatial  infinity. In particular,  setting
 \begin{equation}
 \label{II6}
 \phi = \tau + \chi(\tau, \xi, \theta, \varpi),
 \end{equation}
 where $(\theta, \varpi)$ are angular coordinates of $d\Omega^2$, we require that: (i) the perturbation $\chi$ vanishes at spatial infinity, that is \cite{BS11}, 
 \begin{equation}
 \label{II6.a}
 \chi(\tau, \xi, \theta, \varpi) \rightarrow 0,
 \end{equation}
 as $\xi \rightarrow 0$ (or $ r \rightarrow \infty$). (ii) At the universal horizon, $\chi$ diverges no faster than $|\xi - \xi_{UH}|^{-1}$, that is
  \begin{equation}
 \label{II6.b}
 \chi(\tau, \xi, \theta, \varpi) \rightarrow \left|\xi - \xi_{UH}\right|^{\gamma},
 \end{equation}
as $\xi \rightarrow \xi_{UH}$ ($\equiv 1/r_{UH}$),  where $\gamma \ge -1$.

For  static perturbations, we have \footnote{Since the speed of the khronon is infinitely large, even we consider time-dependent perturbations,
$\chi(\tau, \xi, \theta, \varpi) = T(\tau) \chi(\xi) Y(\theta,\varpi)$, we find that   the resulting field equations of  the linear perturbations do not depend on
$T(\tau)$, that is, $T(\tau)$ now is a undetermined function of $\tau$. This is due to  the instantaneous propagation of the  khronon.}
\begin{equation}
\label{II6.c}
\chi(\tau, \xi, \theta, \varpi) = \chi(\xi) Y(\theta,\varpi),
\end{equation}
where $Y(\theta,\varpi)$ satisfies the equation,
 \begin{equation}
 \label{4}
\Delta_{S^2}Y\equiv \frac{1}{\sin\theta}\frac{\partial}{\partial\theta}\left(\sin\theta\frac{\partial
Y}{\partial\theta}\right)+\frac{1}{\sin^2\theta}\frac{\partial^2Y}{\partial\varpi^2}=-L^2Y.
 \end{equation}
Here, $\Delta_{S^2}$ is the Laplacian operator on the unit two-sphere, and
$L^2 \equiv l(l+1)$ with $l$ being  a non-negative integer. Since it is known that there are no spherically symmetric hairs, hereafter we shall consider non-spherical modes, i.e. modes with positive $l$: $(l = 1, 2, \cdots )$. 

 To study the static perturbations further, in the following let us consider the three backgrounds, Schwarzschild, Schwarzschild anti-de Sitter, and
Reissner-Nordstr\"om, separately.

\subsection{The Schwarzschild Background}

 In this case, from (\ref{2.10}) and (\ref{2.15}) we have \cite{LGSW},
 \begin{equation}
 \label{4.a}
 r_o = \frac{3^{3/4}}{4}r_s, \;\;\; r_{UH} = \frac{3}{4} r_s,
 \end{equation}
where $r_s \equiv 2M$.  Then, choosing $r_s  = 1$ by redefinition of scales, 
we find that the linear static perturbations satisfy the field equation,
 \begin{eqnarray}
 \label{II8}
&&  2\xi^2\left(1- \xi + r_o^4\xi^4\right) H'' + \xi\left(8 - 9 \xi + 12 r_o^4\xi^4\right)H' \nonumber\\
&&  + 2 (2 - L^2 - 3\xi)H = 0,
 \end{eqnarray}
 where
 \begin{eqnarray}
 \label{II9}
 H \equiv \left(1 - \xi + r_o^4\xi^4\right)\left[2\xi^2\left(1 - \xi + r_o^4 \xi^4\right)\chi''\right.\nonumber\\
  \left.- 3\xi^2\left(1- 4r_o^4\xi^3\right)\chi' - 2 L^2\chi\right].
 \end{eqnarray}

At the spatial  infinity, assuming a linear combination of power-law solutions, and then from Eq.(\ref{II8}) we  obtain
 \begin{equation}
 \label{II10}
 \chi \simeq \chi_1\xi^{l+1}+\chi_2\xi^{l-1}+\chi_3\xi^{-l-2}+\chi_4\xi^{-l},
 \end{equation}
 where $\chi_i$'s are the integration constants. Then, the asymptotic condition
 (\ref{II6.a}) requires
 \begin{eqnarray}
 \label{II10.a}
&& (i) \; \chi_3 = \chi_4 = 0,\; (l \ge 2), \nonumber\\
&&   (ii) \; \chi_2 = \chi_3 = \chi_4 = 0, \; (l = 1).
 \end{eqnarray}
Then, from Eq.(\ref{II9}) we find that,
\begin{eqnarray}
\label{3.1}
H = 4\chi_2(1-2l)\xi^{l-1}  + {\cal{O}}\left(\xi^{l}\right),
\end{eqnarray}

 On the other hand, at the universal horizon, Eq.(\ref{II8}) has the solution,
 \begin{eqnarray}
 \label{II11}
 \chi&\simeq&C_1\left|\xi_{UH} - \xi\right|^{\alpha_{+}} +C_2\left|\xi_{UH} - \xi\right|^{\alpha_{+} - 1}\nonumber\\
 &&+C_3\left|\xi_{UH} - \xi\right|^{\alpha_{-}} +C_4\left|\xi_{UH} - \xi\right|^{\alpha_{-}-1}, ~~~~
 \end{eqnarray}
 where $C_i$'s are integration constants, and
 \begin{eqnarray}
 \label{II11.a}
 \alpha_{\pm} \equiv -1\pm \sqrt{1+\frac{l(l+1)}{2}}.
  \end{eqnarray}
Then, the regularity condition (\ref{II6.b}) requires that
\begin{eqnarray}
\label{II11.b}
&& (i) \; C_3 = C_4 = 0,\; (l \ge 2), \nonumber\\
&&   (ii) \; C_2 = C_3 = C_4 = 0, \; (l = 1).
\end{eqnarray}
Since
$$
1 - \xi + r_o^4\xi^4 \simeq {\cal{O}}\left(\left(\xi_{UH} - \xi\right)^{2}\right),
$$
as $ \xi \to \xi_{UH}$, from Eq.(\ref{II9}) we find that
\begin{eqnarray}
\label{3.2}
H &=&   C_2 \times {\cal{O}}\left(\left(\xi_{UH} - \xi\right)^{\alpha_{+}}\right) + {\cal{O}}\left(\left(\xi_{UH} - \xi\right)^{\alpha_{+}+1}\right)\nonumber\\
&\simeq& 0,
\end{eqnarray}
as $ \xi \to \xi_{UH}$. Combining Eqs.(\ref{3.1}) and (\ref{3.2}) we obtain
\begin{equation}
\label{3.3}
H(0) = H\left(\xi_{UH}\right) = 0.
\end{equation}
With these boundary conditions, we shall first show that Eq.(\ref{II8}) does not have nontrivial solution. To this goal, let us first rewrite (\ref{II8}) in the form,
\begin{equation}
\label{3.4}
Y''  - \frac{A_l(\xi)}{B(\xi)} Y = 0,
\end{equation}
where
\begin{eqnarray}
\label{3.4}
Y(\xi) &\equiv& \xi^2\sqrt{3(\xi_{UH} - \xi)}\left(3\xi^3 + 8\xi + 16\right)^{1/4} H(\xi),\nonumber\\
A_l(\xi) &\equiv& 486\xi^6 + 1296 \xi^5 + \xi^2\left(2592\xi^2 -1152 \xi + 768l^2\right) \nonumber\\
&& + 768(l-1)\xi^2 + 2048l(l+1)(\xi+2),\nonumber\\
B(\xi) &\equiv& 9\xi^2(\xi_{UH} - \xi)^2\left(3\xi^3 + 8\xi + 16\right)^2.
\end{eqnarray}
From the above expressions it can be easily seen that 
\begin{equation}
\label{3.5}
Y(0) = Y\left(\xi_{UH}\right) = 0.
\end{equation}
On the other hand, in the range $\xi \in\left(0, \xi_{UH}\right)$, both of the functions $A_l(\xi)$ and $B(\xi)$ are non-negative. Then, Eq.(\ref{3.4})
has the only trivial solution $Y(\xi) = 0$ with the two boundary conditions (\ref{3.5}). As a result, Eq.(\ref{II8}) with the boundary condition (\ref{3.3}) has the unique  solution
\begin{equation}
\label{3.6}
H(\xi) = 0.
\end{equation}
Then,  Eq.(\ref{II9}) reduces to,
 \begin{eqnarray}
 \label{3.7}
  2\xi^2\left(1 - \xi + r_o^4 \xi^4\right)\chi''
  &-& 3\xi^2\left(1- 4r_o^4\xi^3\right)\chi' \nonumber\\
  &-&  2 L^2\chi = 0,
 \end{eqnarray}
where the boundary conditions for $\chi$ are
\begin{equation}
\label{3.8}
\chi(0) = \chi\left(\xi_{UH}\right) = 0,
\end{equation}
as can be seen from Eqs.(\ref{II10})-(\ref{II11.b}). Following what we did for the function $H(\xi)$, it can be shown that with these boundary conditions
Eq.(\ref{3.7}) has the unique solution,
\begin{equation}
\label{3.6}
\chi(\xi) = 0.
\end{equation}
That is, in the current case there are no static hairs even when the speed of the khronon is infinitely large.  We obtain the same conclusion  by using
a different approach, {\em the asymptotic uniform approximation method}, as presented in Appendix A.

 \subsection{The Schwarzschild anti-de Sitter Background}

In the Schwarzschild anti-de Sitter background, we have $
F(\xi)=1-2M\xi-\frac{\Lambda}{3\xi^2}$, where $\Lambda$ is the
cosmological constant and here we assume it to be negative.
 On the other hand, we also have \cite{LGSW},
 \begin{eqnarray}
 \label{III4}
U^2&=& 1-2M\xi-\frac{\Lambda}{3\xi^2}+r_o^4\xi^4\nonumber\\
&=&r_o^4\left(1-\frac{\xi_{UH}}{\xi}\right)^2\Bigg(\xi^4+2\xi_{UH}\xi^3+3\xi_{UH}^2\xi^2\nonumber\\
&& -2\frac{1-3r_o^4\xi_{UH}^4}{3r_o^4\xi_{UH}}\xi-\frac{1-3r_o^4\xi_{UH}^4}{3r_o^4}\Bigg),
 \end{eqnarray}
 where the cosmological constant $\Lambda$, mass of black hole $M$ and the {\ae}ther field parameter  $r_o$ are
 given by, respectively,
 \begin{eqnarray}
 \label{III5}
\Lambda&=&\xi_{KH}^2\xi_{UH}^2\frac{4\xi_{KH}-3\xi_{UH}}{2\xi_{KH}^3-\xi_{UH}^3},\nonumber\\
M&=&\frac{3\xi_{KH}^2-2\xi_{UH}^2}{6\xi_{KH}^3-3\xi{UH}^3},\nonumber\\
r_o^4&=&\frac{\left(\xi_{KH}-\xi_{UH}\right)^2\left(2\xi_{KH}+\xi_{UH}\right)}{3\xi_{UH}^4\left(\xi_{UH}^3-2\xi_{KH}^3\right)},
 \end{eqnarray}
where $\xi_{KH}$ and $\xi_{UH}$ denote the locations of the  killing and universal
horizons, respectively, so they satisfy the relations
$F(\xi_{KH})=U(\xi_{UH})=0$. As we all know, mass of black hole
should be positive definite and $\Lambda<0$ for Anti-de Sitter
spacetime. Therefore, we have

\begin{equation}
\label{rs}
\frac{3}{4}<\frac{\xi_{KH}}{\xi_{UH}} <\frac{1}{2^{1/3}}.
\end{equation}
 The metric will reduce to
the Schwarzschild case when $\xi_{KH}/\xi_{UH}=3/4$. Then, the perturbation
equation for $\chi(\xi)$ becomes,
 \begin{eqnarray}
 \label{III7}
 &&\xi^2  H'' \left[\Lambda  -3 \xi ^2 \left(-2 M \xi +\xi ^4
   r_o^4+1\right)\right]\nonumber\\
   &&+\xi H'  \left[\Lambda -3\xi ^2 \left(2-5 M
   \xi +4r_o^4\xi^4\right)\right]\nonumber\\
   &&+ H \left[3\xi^2\left(L^2+M\xi+4r_o^4\xi^4\right)-4\Lambda\right]=
   0,
 \end{eqnarray}
 where
 \begin{eqnarray}
 \label{III8}
 H &\equiv& \xi^{-2}\left[\Lambda -3 \xi ^2 \left(-2 M \xi +\xi ^4 r_o^4+1\right)\right] \left\{3 L^2  \xi  \chi\right.\nonumber\\
 &&-3\chi'\left(\Lambda -3 M \xi ^3+6 \xi ^6 r_o^4\right)\nonumber\\
 &&\left.+\xi  \chi'' \left[\Lambda -3 \xi^2 \left(-2 M \xi +\xi ^4 r_o^4+1\right)\right]\right\}.  ~~~~~
 \end{eqnarray}
Then, near the spatial  infinity, since $\Lambda\not=0$, the solution   is given by
 \begin{equation}
 \label{III9}
 \chi \simeq \chi_1\xi^5+\chi_4\xi^4+\chi_3\xi+\chi_4,
 \end{equation}
where again $\chi_i$'s are integration constants. Near the universal horizon, on the other hand, the solution takes the form,  
 \begin{eqnarray}
 \label{III10}
 \chi&\simeq &C_1\left(\xi_{UH}-\xi\right)^{\beta -2}  +C_2\left(\xi_{UH}-\xi\right)^{\beta-1}\nonumber\\
 &&+C_3\left(\xi_{UH}-\xi\right)^{-\beta -2} +C_4\left(\xi_{UH}-\xi\right)^{-\beta-1},~~~~~~~
 \end{eqnarray}
  where
 \begin{eqnarray}
 \label{III10aa}
\beta \equiv
\sqrt{\frac{(2+l+l^2)\xi_{UH}^3-2(l^2+l-4)\xi_{KH}^3-9\xi_{KH}^2\xi_{UH}}{2\xi_{UH}^3-9\xi_{KH}^2\xi_{UH}+9\xi_{KH}^3}}.\nonumber\\
 \end{eqnarray}
Then, the boundary conditions (\ref{II6.a}) and (\ref{II6.b})
require that
 \begin{equation}
 \label{III10a}
 \chi_4 =  C_3=C_4=0.
 \end{equation}
 
\begin{figure}[tbp]
\centering
\includegraphics[width=8cm]{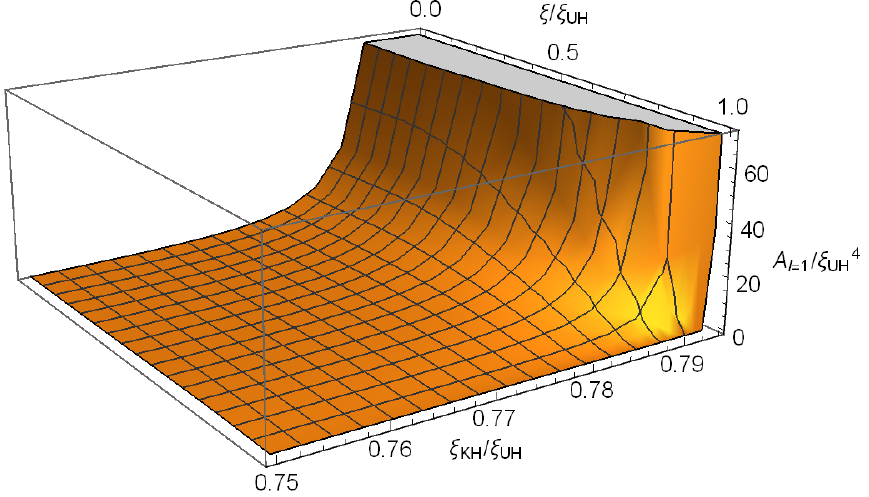}
\caption{The function $A_{l=1}(\xi)/\xi_{UH}^4$ vs $\xi_{KH}/\xi_{UH}$ and
$\xi/\xi_{UH}$, where $3/4 < \xi_{KH}/\xi_{UH}< 2^{-1/3}$ and
$0 < \xi/\xi_{UH} < 1$.  It is clear that $A_{l=1}(\xi)$ is always positively definite.} \label{figA}
\end{figure}

Inserting Eqs.(\ref{III9}) and (\ref{III10a}) into Eq.(\ref{III8}), we find that at the spatial infinity ($\xi \simeq 0$) we have 
 \begin{eqnarray}
 \label{III10c}
 H(\xi)=-3\chi_3\xi_{UH}^4\xi_{KH}^4\frac{\left(4\xi_{KH}-3\xi_{UH}\right)^2}{\left(\xi_{UH}^3-2\xi_{KH}^3\right)^2}+
 {\cal{ O}}\left(\xi^2\right). ~~~~
 \end{eqnarray}
 On the other hand, near the universal horizon $\xi \simeq \xi_{UH}$, we have
  \begin{eqnarray}
 \label{III10d}
 H(\xi)\simeq (\xi-\xi_{UH})^\alpha, 
 \end{eqnarray}
where $\alpha>2$. Following what we did in the Schwarzschild case, we also
rewrite Eq.(\ref{III8}) in the form, 
 \begin{eqnarray}
 \label{III11}
 Y''-\frac{A_l(\xi)}{B(\xi)}Y=0
 \end{eqnarray}
but now with  
 \begin{eqnarray}
 \label{III12}
 Y(\xi)&\equiv&\xi^{1/2}\left[3\xi^2\left(1-2M\xi+r_o^4\xi^4\right)-\Lambda\right]^{1/4}H(\xi),\nonumber\\
 A_l(\xi)&\equiv&4L^2\left[3\xi^2\left(1-2M\xi+r_o^4\xi^4\right)-\Lambda\right]\nonumber\\
 &&+5\Lambda^2-6\Lambda\xi^2\left(3-7M\xi+11r_o^4\xi^4\right)\nonumber\\
 &&-9\xi^6\left[M^2+20Mr_o^4\xi^3-4r_o^4\xi^2\left(3+2r_o^4\xi^4\right)\right],\nonumber\\
 B(\xi)&\equiv&\frac{4}{3}\xi^2\left[3\xi^2\left(1-2M\xi+r_o^4\xi^4\right)-\Lambda\right]^2.
 \end{eqnarray}
From the expression of $A_l$ one can see that it takes its minimal value when $l = 1$, that is,  $A_{l \ge 2} >  A_{l=1}$, considering the fact that $\Lambda < 0$ and $L^2=l(l+1)$.
Thus, to prove that $A_l$ is no-negative, it is sufficient to prove that $A_{l =1}$ is non-negative. 
Fig.(\ref{figA}) shows that $A_{l=1}(\xi)$ is indeed non-negative in the region $\xi \in (0, \xi_{UH})$ with the constraints of Eq.(\ref{rs}).

On the other hand, from Eqs.(\ref{III10c}), (\ref{III10d}) and (\ref{III12}) we find that at  the boundaries $\xi = 0$ and $\xi = \xi_{UH}$, the function $Y$ satisfies
  \begin{eqnarray}
 \label{III13}
Y(0)=Y(\xi_{UH})=0.
 \end{eqnarray}
 Then, similar to the Schwarzschild case, since the functions $A_l$ and $B$ defined by Eq.(\ref{III12}) are non-negative for $\xi\in(0, \xi_{UH})$,  we must have
 $Y(\xi) = 0$. Then, from Eq.(\ref{III12}) we find that
 \begin{eqnarray}
 \label{III14}
H(\xi) = 0, 
 \end{eqnarray}
which means that $\chi_3 = 0$ and 
 \begin{eqnarray}
 \label{III15}
\xi\left[\Lambda-3\xi^2\left(1-2M\xi+r_o^4\xi^4\right)\right]\chi''\nonumber\\
-3\left(\Lambda-3M\xi^3+6r_o^4\xi^6\right)\chi'+3L^2\xi\chi=0.
 \end{eqnarray}
Setting
 \begin{equation}
 \label{III15a}
 \chi(\xi) = \frac{\xi^{3/2}\hat\chi(\xi)}{\left[3\xi^2\left(1-2M\xi+r_o^4\xi^4\right)-\Lambda\right]^{1/4}}, 
 \end{equation}
we find that Eq.(\ref{III15}) can be cast exactly  in the same form of
Eq.(\ref{III11}) in terms of $\hat\chi$, where $A_l$ and $B$ are also given by Eq.(\ref{III12}). From the boundary
conditions $\chi(0)=\chi(\xi_{UH})=0 $ we find that
 \begin{eqnarray}
 \label{III17}
\hat\chi(0) = 0 = \hat\chi(\xi_{UH}).
 \end{eqnarray}
Again, since $A_l$ and $B$ defined by Eq.(\ref{III12}) are all
no-negative for  $\xi\in(0, \xi_{UH})$, we must have $\hat\chi(\xi)
= 0$. Then, we conclude that
 \begin{equation}
 \label{III18}
 \chi(\xi) = 0.
 \end{equation}
Thus, in the present case long static hairs do not exist either.

\subsection{The Reissner-Nordstr\"om Background}

Finally, let's consider the  static hairs in the Reissner-Nordstr\"om background, for which we have  $F=1-2M\xi+Q^2\xi^2$, and
 \begin{eqnarray}
 \label{IV1}
U^2&=& 1-2M\xi+Q^2\xi^2+r_o^4\xi^4\nonumber\\
&=&r_o^4\left(\xi-\xi_{UH}\right)^2\left(\xi^2+2\xi_{UH}\xi+\frac{1}{r_o^4\xi_{UH}^2}\right),  ~~~~~
 \end{eqnarray}
but now $r_o$ and $\xi_{UH}$ are given via the relations,
 \begin{eqnarray}
 \label{IV2}
Q^2=\frac{1-3r_o^4\xi_{UH}^4}{\xi_{UH}^2},~~~~~M=\frac{1-r_o^4\xi_{UH}^4}{\xi_{UH}}.
 \end{eqnarray}
Clearly, since $Q^2 \ge 0$, we must have  $0\le r_o^4\le\frac{1}{3\xi_{UH}^4}$. When $r_o = 0$, the charged black hole becomes
extreme. 

\begin{figure}[tbp]
\centering
\includegraphics[width=8cm]{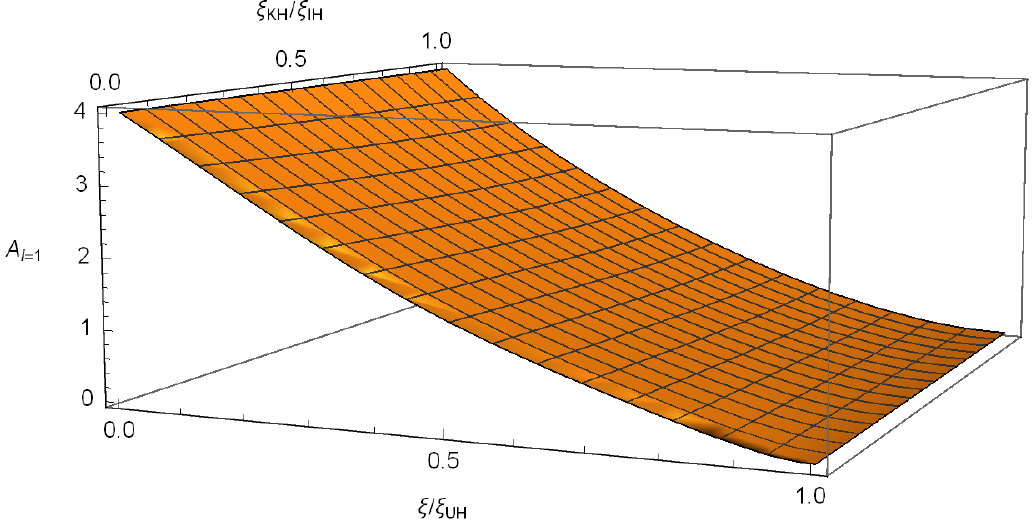}
\caption{The function $A_{l=1}(\xi)$ vs $\xi_{KH}/\xi_{IH}$ and $\xi/\xi_{UH}$,
where $0\le\xi_{KH}/\xi_{IH}\le1$ and $0\le\xi/\xi_{UH}\le1$.} \label{figB}
\end{figure}

In the general case, it can be shown that the   perturbation satisfies the equation,
 \begin{eqnarray}
 \label{IV3}
&&\chi^{(4)}+\left(\frac{8}{\xi }+\frac{12 U'}{U}\right)\chi^{(3)}+\left(-2L^2 +36 \xi^2 U'^2\right.\nonumber\\
&&\left.+3 \xi U\left(3 \xi  U''+20 U'\right)+12 U^2\right)\frac{\chi''}{\xi ^2 U^2}+\left\{-8 \xi  U'\left(\lambda\right.\right.\nonumber\\
&& \left.-3 \xi^2U'^2\right)+U\left(-4 L^2 +84 \xi ^2 U'^2+27 \xi^3 U' U''\right)\nonumber\\
&&\left.+3 \xi  U^2 \left[12 U'+\xi  \left(\xi  U^{(3)}+8 U''\right)\right]\right\}\frac{\chi'}{\xi^3 U^3}\nonumber\\
&&+\left(-2 c_o \xi ^2 Q^2+L^2 -3 \xi ^2 U'^2\right.\nonumber\\
&&\left.-\xi  U\left(\xi U''+10 U'\right)-2 U^2\right)\frac{L^2}{\xi^4U^4}\chi= 0,
 \end{eqnarray}
where $c_o \equiv c_3/(c_2+c_3)$ and $\chi^{(3)} \equiv d^3\chi(\xi)/d\xi^3$, etc. For non-zero values of $c_3$, we have not succeeded in 
proving that there are no long hairs. However, when $c_3 = 0$, we find that the proof can be followed in a similar way as we did in the last two cases. In fact, when 
$c_3 = 0$, the perturbation of $\chi$  becomes
 \begin{eqnarray}
 \label{IV4a}
&&\xi^2\left[2-4M\xi+2Q^2\xi^2+2r_o^4\xi^4\right]H''\nonumber\\
&&+\xi\left[8-18M\xi+10Q^2\xi^2+12r_o^4\xi^4\right]H'\nonumber\\
&&+\left[4-2l(l+1)+3\xi(2Q^2\xi-4M)\right]H=0,
 \end{eqnarray}
or
 \begin{eqnarray}
 \label{IV4b}
Y''-\frac{A_l(\xi)}{B(\xi)}Y=0,
 \end{eqnarray}
where
 \begin{eqnarray}
 \label{IV5}
H(\xi)&\equiv&(4M\xi-2-2Q^2\xi^2-2r_o^4\xi^4)\nonumber\\
&&\times\left[\xi\left(2-4M\xi+2Q^2\xi^2+2r_o^4\xi^4\right)\chi''\right.\nonumber\\
&&\left.+3\xi\left(2Q^2\xi+4r_o^4\xi^3-2M\right)\chi'-2l(l+1)\chi\right],\nonumber\\
\\
 \label{IVt}
Y(\xi)&\equiv&\xi^2\left(1-2M\xi+Q^2\xi^2+r_o^4\xi^4\right)^{1/4}H(\xi),\\
 \label{IVx}
A_l(\xi)&\equiv&l(l+1)\left(4-8M\xi+4Q^2\xi^2+4r_o^4\xi^4\right)\nonumber\\
&&+(6Q^2-3M^2)\xi^2-6MQ^2\xi^3\nonumber\\
&&+3(Q^4+12r_o^4)\xi^4-60Mr_o^4\xi^5\nonumber\\
&&+30Q^2r_o^4\xi^6+24r_o^6\xi^8,\nonumber\\
B(\xi)&\equiv&4\xi^2\left(1-2M\xi+Q^2\xi^2+r_o^4\xi^4\right)^2.
 \end{eqnarray}
In this case, it can be shown that the following relations hold among the parameters of the RN solutions and the locations of various horizons, 
 \begin{eqnarray}
 \label{IVy1}
 M&=&\frac{1-r_o^4\xi_{UH}^4}{\xi_{UH}},~~Q^2=\frac{1-3r_o^4\xi_{UH}^4}{\xi_{UH}^2},\nonumber\\
 r_o^4&=&\frac{2b-3-3b^2+(b+1)\Delta}{8b\xi_{UH}^4},\nonumber\\
 \Delta&=& \sqrt{9b^2-14b+9},
 \end{eqnarray}
where $b\equiv \xi_{KH}/\xi_{IH}$, and $\xi_{KH}$ and $\xi_{IH}$ are the positions of 
the  Killing and inner horizons,  respectively. Because $0\le b\le1$ and $\xi_{KH}\ge\xi_{UH}\ge0$, it can be shown that $A_l(\xi)$ and 
$B(\xi)$ defined above  are positively defined for $\xi \in (0, \xi_{UH})$. The proof of the positivity of $A_l(\xi)$ can follow a similar procedure as given above. In particular, the  positivity of $A_{l = 1}(\xi)$ is shown in Fig.\ref{figB}.

Now, let's consider the boundary conditions at the spatial infinity and the universal horizon. First, near the spatial infinity, $\chi$ is given by,
 \begin{equation}
 \label{IV5a}
 \chi\simeq \chi_1\xi^{l+1}+\chi_2\xi^{l-1}+\chi_3\xi^{-l-2}+\chi_4\xi^{-l}.
 \end{equation}
Then,  the boundary conditions at the spatial  infinity  require
 \begin{eqnarray}
 \label{IV5b}
(i)~&&\chi_3=\chi_4=0~~(l=1)\nonumber\\
(ii)~&&\chi_2=\chi_3=\chi_4=0~~(l\ge2),
 \end{eqnarray}
from which we find that near $\xi \simeq 0$ we have
 \begin{equation}
 \label{IV5b}
H = \chi_2(1-2l)\xi^{l-1}+{\cal O}(\xi^l).
 \end{equation}
 On the other hand, at the universal horizon, $\chi$ takes the form,
 \begin{eqnarray}
 \label{IV6a}
 \chi&\simeq &C_1\left(\xi_{UH}-\xi\right)^{\gamma-1}+C_2\left(\xi_{UH}-\xi\right)^{\gamma-2}\nonumber\\
 &&+C_3\left(\xi_{UH}-\xi\right)^{-\gamma-1}+C_4\left(\xi_{UH}-\xi\right)^{-\gamma-2},\nonumber\\
 \end{eqnarray}
 where
 \begin{eqnarray}
 \label{IV7a}
 \gamma&=&\sqrt{\frac{8bl(l+1)+3b\Delta+\Delta(3-\Delta)}{\Delta(\Delta-3b-3)}}.
 \end{eqnarray}
  Thus, the boundary conditions at the universal horizon require
 \begin{eqnarray}
 \label{IV7b}
 (i)&&~C_2=C_3=C_4=0~~(l=1),\nonumber\\
 (ii)&&~C_3=C_4=0~~(l\ge2),
 \end{eqnarray}
 from which we find that  $H(\xi_{UH}) = 0$. Then, from Eq.(\ref{IVt}) we find that   $Y(0)=Y(\xi_{UH})=0$. As mentioned above, 
  $A_l$ and $B$ are all positively defined, so Eq.(\ref{IV4b}) has only zero solution. Then, we find that
$H(\xi) = 0$. Then, from Eq.(\ref{IV5}) we find that
\begin{eqnarray}
\label{IV7c}
&&  \xi\big(2-4M\xi+2Q^2\xi^2 +2r_o^4\xi^4\big)\chi'' \nonumber\\
&& +3\xi\left(2Q^2\xi+4r_o^4\xi^3-2M\right)\chi'-2L^2\chi = 0.
\end{eqnarray}
Then, setting 
 \begin{eqnarray}
 \label{IV8}
\chi(\xi)\equiv\frac{\hat\chi(\xi)}{\left(2-4M\xi+2Q^2\xi^2+2r_o^4\xi^4\right)^{3/4}},
 \end{eqnarray}
we find that $\hat{\chi}(\xi)$ satisfies the same equation (\ref{IV4b})  with the same functions $A_l$ and $B$.
Then, with the boundary conditions $\hat{\chi}(0) =  \hat{\chi}(\xi_{UH}) = 0$, we find that $\hat\chi(\xi) = 0$, so that  
 \begin{eqnarray}
 \label{IV7c}
\chi(\xi) = 0,
 \end{eqnarray}
for $\xi \in (0, \xi_{UH})$.  Thus, the the RN case, long hairs do not exist either in the case $c_3 = 0$.

For non-zero values of $c_3$, we still expect that the forth order equation (\ref{IV3}) does not allow for long hairs in the RN background. One of the reasons why we expect so is that we do not have enough number of shooting parameters as we shall see now. We first redefine the scale so that $\xi_{UH}=1$. Then, near the spatial  infinity, $\chi$ is given by,
 \begin{equation}
 \chi\simeq \chi_1\xi^{l+1}+\chi_2\xi^{l-1}+\chi_3\xi^{-l-2}+\chi_4\xi^{-l},
 \end{equation}
while near the universal horizon,  it takes the form,
 \begin{eqnarray}
 \chi&\simeq &C_1\left(1-\xi\right)^{-\frac{3}{2}+\frac{\gamma_-}{2}}+C_2\left(1-\xi\right)^{-\frac{3}{2}+\frac{\gamma_+}{2}}\nonumber\\
 &&+C_3\left(1-\xi\right)^{-\frac{3}{2}-\frac{\gamma_-}{2}}+C_4\left(1-\xi\right)^{-\frac{3}{2}-\frac{\gamma_+}{2}}
 \end{eqnarray}
 where
 \begin{eqnarray}
 \gamma_\pm&=&\left\{\frac{4 l^2+4 l+15 r_o^4+5}{3 r_o^4+1}\pm\frac{4}{3 r_o^4+1}\right.\nonumber\\
 &&\times\left[l^2 \left(-6 c_o r_o^4+2 c_o+3 r_o^4+1\right)+\left(3 r_o^4+1\right)^2\right.\nonumber\\
 &&\left.\left.+l \left(-6 c_o r_o^4+2 c_o+3
   r_o^4+1\right)\right]^{1/2}\right\}^{1/2}.
 \end{eqnarray}
  Thus, the boundary conditions require
 \begin{equation}
\chi_3=\chi_4=C_3=C_4=0,
 \end{equation}
 leaving the four coefficients ($\chi_1$, $\chi_2$, $C_1$, $C_2$) as free parameters. Among the four free parameters, one of them (or one linear combination of them) corresponds to an overall normalization and can be set to $1$, for example. We then have only three free parameters. We can then numerically solve the forth-order equation (\ref{IV3}) from both boundaries and try to match one solution from one side to the other solution from the other side somewhere in the middle, say at $\xi=\xi_{UH}/2$. Since we are dealing with a forth-order equation, the matching condition consists of the continuities of four quantities, i.e. ($\chi$, $\chi'$, $\chi''$, $\chi^{(3)}$). However, as we have seen, there are only three free parameters that we can use as shooting parameters. Thus we expect that there should be no solution to this shooting problem. This argument is rather generic but, unfortunately, does not exclude the possibility that the matching condition may be satisfied for special values of the  parameters $c_{1,\cdots,4}$. Therefore, the proof of no-hair for the RN background with non-vanishing values of $c_3$ still remains an open question.

\section{Large $c_{\phi}$ Expansions }
\renewcommand{\theequation}{4.\arabic{equation}} \setcounter{equation}{0}

In the previous section, we considered only the case where   $c_{\phi} = \infty$ (or $c_{14}=0$). In this section, we shall study the case where $c_{\phi}$ is large but finite. For simplicity, in this section we shall restrict ourselves only to the  Schwarzschild background. Then,  the functions  $U$ and $V$   satisfy the equations,
 \begin{eqnarray}
 \label{new1}
&& \frac{U''(\xi)}{U(\xi)}-c_{\phi}^2\frac{V''(\xi)}{V(\xi)}+\frac{2c_\phi^2}{\xi^2} = 0,\\
&& U(\xi)^2-V(\xi)^2-1+\xi = 0.
 \end{eqnarray}
Setting
 \begin{eqnarray}
 \label{new2}
V(\xi)=-r_o^2\xi^2
+\epsilon^2 V_2(\xi)+{\cal
O}(\epsilon^4),
 \end{eqnarray}
where $\epsilon \equiv {1}/{c_\phi}$,
from
Eq.(\ref{new1}) we find that,
 \begin{eqnarray}
 \label{new3}
V''_2-\frac{2}{\xi^2}V_2
+ \frac{3\sqrt{3}\xi^2(27\xi^3+108\xi^2+288\xi+128)}{24\left(\xi-\frac{4}{3}\right)\left(3\xi^2+8\xi+16\right)^2} = 0.\nonumber\\
 \end{eqnarray}
It is interesting to note that the above equation is singular at $\xi = 4/3$. The general solution of the above equation is given by,
 \begin{eqnarray}
 \label{new4}
V_2(\xi)&=&C_1\xi^2+\frac{C_2}{\xi}+\frac{448-27\xi^3}{864\sqrt{6}\xi}\arctan\left(\frac{3\xi+4}{4\sqrt{2}}\right)\nonumber\\
&&-\frac{1}{216\sqrt{3}\xi}\Bigg\{304-9\xi(3\xi^2+3\xi+4)\nonumber\\
&&+ \left(27\xi^3 + 32\ln3\right)\ln\left(3\xi^2+8\xi+16\right)\nonumber\\
&& + 27\left[\xi^3 - \left(\frac{4}{3}\right)^3\right]\ln\left(3\left|\xi - \frac{4}{3}\right|\right)\Bigg\},
 \end{eqnarray}
where $C_1$ and $C_2$ are constants. Note that $V_2(\xi)$ given above is well-defined at $\xi = 4/3$, although Eq.(\ref{new3}) is singular there.
The boundary condition $U(\xi=0)=1$ leads to
 \begin{eqnarray}
 \label{new5}
C_2=\frac{-1}{27\sqrt{3}}\left[7\sqrt{2}\arctan\left(\frac{1}{\sqrt{2}}\right)-2\ln(9)-38\right].
 \end{eqnarray}
Then, $G(\xi)$ defined in (\ref{2.13}) is 
 \begin{eqnarray}
 \label{new6}
G(\xi) &=& U(\xi)^2\nonumber\\
& = & 1-\xi+\frac{27\xi^4}{256}\nonumber\\
&&+\frac{\epsilon^2\xi}{4608}\left\{27\left[\xi^3-\left(\frac{4}{3}\right)^3\right]\ln\left(3\left|\xi - \frac{4}{3}\right|\right)\right.\nonumber\\
&& + \sqrt{2}\left(27\xi^3-448\right)\arctan\left(\frac{4+3\xi}{4\sqrt{2}}\right)\nonumber\\
&&-72\xi\left[4+3\xi(1+\xi+8\sqrt{3}C_1\xi)\right]\nonumber\\
&&+(32+27\xi^3)\ln\left(16+8\xi+3\xi^2\right) \nonumber\\
&& +56\sqrt{2}\text{arccot}\left(\sqrt{2}\right)\Bigg\}+{\cal
O}(\epsilon^4).
 \end{eqnarray}

When $c_\phi$ is not infinitely large, the universal horizon $\xi_{UH}$
is different from  the sound horizon $\xi_c$, which now satisfies the
equations \cite{BS11},
 \begin{eqnarray}
 \label{new6a}
G(\xi_c)&=&\frac{1-\xi_c}{1-c_\phi^2},\nonumber\\
\left.\frac{dG}{d\xi}\right|_{\xi=\xi_c}&=&(1-c_\phi^2)^{-1}\nonumber\\
&&\times\left[-1+c_\phi\sqrt{1-\frac{8c_\phi^2(1-c_\phi^2)G^2(\xi_c)}{\xi_c^2}}\right].\nonumber\\
 \end{eqnarray}
 Expanding them
  in terms of  $\epsilon$,  we find that
 \begin{eqnarray}
 \label{new7a}
G(\xi_c)&=&-\epsilon^2(1-\xi_c)+{\cal
O}(\epsilon^4),\nonumber\\
\left.\frac{dG}{d\xi}\right|_{\xi=\xi_c}&=&-\epsilon\sqrt{9-\frac{16}{\xi_c}+\frac{8}{\xi_c^2}}+\epsilon^2+{\cal
O}(\epsilon^3).\nonumber\\
 \end{eqnarray}

 On the other hand, from Eq.(\ref{new6}) we can see that the term,
 $$
 \left[\xi^3-\left(\frac{4}{3}\right)^3\right]\ln\left(3\left|\xi - \frac{4}{3}\right|\right) \simeq
 \epsilon\ln(\epsilon)F_1 + ...
 $$
 for $\xi_c = 4/3 + \epsilon \xi_1 + ...$, where $\xi_1$ and $F_1$ are finite and non-zero constants. As
 a result, $G(\xi_c)$ contains terms proportional to $ \epsilon^3\ln(\epsilon)$, while $dG(\xi_c)/d\xi$
 contains terms proportional to $ \epsilon^2\ln(\epsilon)$. Then, for Eqs.(\ref{new6}) and
 (\ref{new7a}) to be consistent, the constant  $\xi_c$ must take the form,
 \begin{eqnarray}
 \label{new7}
\xi_c=\frac{4}{3}+ \epsilon\xi_1+ \epsilon^2\ln(\epsilon)\xi_2 +\epsilon^2\xi_3+
...
 \end{eqnarray}
where $\xi_i \; (i = 1, 2, 3)$ are constants. Inserting it into   Eqs.(\ref{new6}) and
 (\ref{new7a}),   we find
 \begin{eqnarray}
 \label{new10}
C_1&=&\frac{1}{64\sqrt{3}}\left[12\ln(32)-20-6\sqrt{2}\arctan(\sqrt{2})\right.\nonumber\\
&&\left.+7\sqrt{2}\arctan\left(\frac{1}{\sqrt{2}}\right)\right],\nonumber\\
\xi_1&=&-\left(\frac{2}{3}\right)^{3/2},~~~~~
\xi_2 = -\frac{4}{27}, \nonumber\\
\xi_3&=&\frac{1}{54}\Bigg[8+7\sqrt{2}\arctan\left(\frac{1}{\sqrt{2}}\right)-7\sqrt{2}\arctan\left(\sqrt{2}\right)\nonumber\\
&&+\ln\left(20736\right)\Bigg].
 \end{eqnarray}

Similarly,   $\xi_{UH}$ should take the form,
$\xi_{UH}={4}/{3}  + \epsilon\xi_\alpha  +\epsilon^2\xi_\beta+
...$
 Then, from Eq.(\ref{new6}), which  is valid up to the second order in $\epsilon$, we find that $U^2(\xi_{UH})=0$ leads to $\xi_\alpha=0$ , 
 but with $\xi_\beta$ being undetermined. Therefore, to determine $\xi_{\beta}$ it is necessary to consider high order expansions. 
 However, to our current purpose we shall leave $\xi_\beta$ as undetermined, so that    $\xi_{UH}$ is given by,
 \begin{eqnarray}
 \label{new11}
\xi_{UH}&=&\frac{4}{3}   +\epsilon^2\xi_\beta+ ...
 \end{eqnarray}

On the other hand, the perturbation equation of the khronon field
is given by \cite{BS11}
 \begin{eqnarray}
 \label{new13}
(\hat A(\xi)\chi'')''-(\hat B_l(\xi)\chi')'+\hat C_l(\xi)\chi=0,
 \end{eqnarray}
where $\hat A(\xi), \hat B_l(\xi)$ and $\hat C_l(\xi)$ are given by Eq.(\ref{new14}) in Appendix B.
Expanding them  in terms of $\epsilon$,  we find that these coefficients take the forms of
Eq.(\ref{new14a}).  Setting
 \begin{equation}
 \label{EXP1}
 \chi =\bar\chi + \epsilon^2\tilde\chi + ...,
 \end{equation}
we find that to the zeroth-order of $\epsilon$, Eq.(\ref{new13}) reduces precisely to the ones given by
Eqs.(\ref{II8}) and (\ref{II9}), while to the second-order, it reads,
 \begin{eqnarray}
 \label{EXP2}
&&\tilde\chi''''+\hat P_3\tilde\chi'''+\hat P_2\tilde\chi''+\hat P_1\tilde\chi'+\hat P_0\tilde\chi+ F(\xi) = 0,
~~~~~~~~~
 \end{eqnarray}
where $ F(\xi)  \equiv \hat Q_4\bar\chi'''' +\hat Q_3\bar\chi'''+\hat Q_2\bar\chi''+\hat
Q_1\bar\chi'+\hat Q_0\bar\chi$, and $\hat P_i$ and $\hat Q_i$ are given by Eq.(\ref{EXP3}) in Appendix B.

 At the infinity ($\xi = 0$), we find
 \begin{eqnarray}
 \label{EXP4}
 V_2&=&\frac{\xi^2}{32\sqrt{3}}\Big\{\ln(64)-7\nonumber\\
 &&\left.+3\sqrt{2}\left[\text{arccot}(\sqrt{2})-\arctan(\sqrt{2})\right]\right\}+{\cal
 O}(\xi^3),\nonumber\\
 \hat P_3&=&\frac{8}{\xi}+{\cal O}(\xi^0),\nonumber\\
 \hat P_2&=&\frac{12-2l^2}{\xi^2}+{\cal O}(\xi^{-1}),\nonumber\\
 \hat P_1&=&-\frac{4l^2}{\xi^3}+{\cal O}(\xi^{-2}),\nonumber\\
 \hat P_0&=&\frac{l^4-2l^2}{\xi^4}+{\cal O}(\xi^{-3}),\nonumber\\
 \hat Q_4&=&-\frac{9\xi^4}{256}\Big\{\ln(64)-4\nonumber\\
 &&\left.+3\sqrt{2}\left[\text{arccot}(\sqrt{2})-\arctan(\sqrt{2})\right]\right\}+{\cal O}(\xi^5),\nonumber\\
 \hat Q_3&=&-\frac{9\xi^3}{16}\Big\{\ln(64)-4\nonumber\\
 &&\left.+3\sqrt{2}\left[\text{arccot}(\sqrt{2})-\arctan(\sqrt{2})\right]\right\}+{\cal O}(\xi^4),\nonumber\\
 \hat Q_2&=&\frac{1}{4}+{\cal O}(\xi),\nonumber\\
 \hat Q_1&=&\frac{1}{\xi}+{\cal O}(\xi^0),\nonumber\\
 \hat Q_0&=&-\frac{3l^2}{256}\Big\{156-7l^2+(l^2-30)\ln(64)\nonumber\\
 &&\left.+3\sqrt{2}(l^2-30)\left[\text{arccot}(\sqrt{2})-\arctan(\sqrt{2})\right]\right\}\nonumber\\
 && +{\cal O}(\xi). 
 \end{eqnarray}
On the other hand, the leading term of $\bar\chi$ in this limit  is  $\bar\chi \simeq \chi_2\xi^{l-1}$ [cf. Eqs.(\ref{II10}) and (\ref{II10.a})]. Substituting
the above expressions  into Eq.(\ref{EXP2}), we get
 \begin{eqnarray}
 \label{EXP5}
 \tilde\chi&\approx& - \frac{\chi_2(l-1)(l+2)}{4l(5l +6)}\xi^{l+1} +\tilde\chi_1\xi^{\frac{1+\sqrt{1+4l^2}}{2}}\nonumber\\
 &&+\tilde\chi_2\xi^{\frac{1-\sqrt{1+4l^2}}{2}} +\tilde\chi_3\xi^{\frac{-3-\sqrt{1+4l^2}}{2}}\nonumber\\
 &&+\tilde\chi_4\xi^{\frac{-3+\sqrt{1+4l^2}}{2}},
 \end{eqnarray}
where   $\tilde\chi_i$ are integration constants. Thus, the asymptotic condition (\ref{II6.a}) requires \footnote{It should be noted that, unlike the case
$c_{\phi} = \infty$, now the $\tilde\chi_4$ term in Eq.(\ref{EXP5}) becomes unbounded for $l = 1$. So, in order to have a finite $\tilde\chi_i$, here we
also set
$\tilde\chi_4 = 0$,  although this will not affect our final conclusions, as to be shown below.},
\begin{equation}
\label{CDa}
\tilde\chi_2 = \tilde\chi_3 = \tilde\chi_4 = 0.
\end{equation}

On the other hand, at the sound horizon we find
 \begin{eqnarray}
 \label{EXP6}
 V_2&\simeq&  -\frac{ 1152}{2304\sqrt{3}}\left(\xi - \frac{4}{3}\right)\ln\left(\xi - \frac{4}{3}\right),\nonumber\\
 \bar\chi &\simeq& C_2\left(\xi - \frac{4}{3}\right)^{\sigma},
 \end{eqnarray}
where we assumed  $\xi - 4/3 \gtrsim  0$, and
\begin{eqnarray}
\label{sigma}
 \sigma \equiv \alpha_{+}-1  &=& - 2 + \sqrt{1 + \frac{l(l+1)}{2}} \nonumber\\
 &=&
\begin{cases}
 < 0, & l = 1,\cr
\ge  0, & l \ge  2. \cr
\end{cases}
\end{eqnarray}
Then,   Eq.(\ref{EXP2})  reduces to
 \begin{eqnarray}
 \label{EXP6}
 \tilde\chi'''' &+& \frac{12}{\xi-4/3}\tilde\chi'''+\frac{36-l^2}{(\xi-4/3)^2}\tilde\chi''\nonumber\\
&+& \frac{24-4l^2}{(\xi-4/3)^3}\tilde\chi' + F(\xi) = 0,
 \end{eqnarray}
 where
 \begin{equation}
 \label{FF}
F(\xi) =  \frac{l^2(l^2-6)\bar\chi}{4(\xi-4/3)^4}+\frac{4\sigma(1+\sigma)(l^2-2\sigma-2\sigma^2)\bar\chi}{27(\xi-4/3)^{6}}.
 \end{equation}
 The general solution of Eq.(\ref{EXP6})  is
 \begin{eqnarray}
 \label{EXP7}
\tilde\chi&\approx&\frac{1}{3(4-3\xi)^2}\left\{\tilde\chi_{a}(\xi-4/3)^{-\sqrt{1+\frac{l^2}{2}}}\right.\nonumber\\
&&+\tilde\chi_{b}(\xi-4/3)^{\sqrt{1+\frac{l^2}{2}}}\nonumber\\
&& +\tilde\chi_{c}(\xi-4/3)^{1-\sqrt{1+\frac{l^2}{2}}}\nonumber\\
&&+\tilde\chi_{d}(\xi-4/3)^{1+\sqrt{1+\frac{l^2}{2}}}\nonumber\\
&&+\hat{C}_{2}  \left(\xi -\frac{4}{3}\right)^{\sigma },
 \end{eqnarray}
 where
 \begin{eqnarray}
 \label{hC2}
 \hat{C}_{2}  \equiv  \frac{16  \sigma  (\sigma +1) \left(2 \sigma (\sigma
+1)-l^2\right) C_2}{l^4-l^2\left(4\sigma^2-4\sigma-2\right)+4\sigma^4-8\sigma^3-4\sigma^2+8\sigma}.\nonumber\\
 \end{eqnarray}

In the studies of the formation of black holes from gravitational collapse in the framework of the Einstein-aether theory \cite{GEJ}, it was found that the
solutions are always regular across the sound horizon. In addition, the regularity of static spherically symmetric solutions also reduces the 2-dimensional
parameter space of black holes to 1-dimensional, whereby the first-law of black hole mechanics is saved \cite{BJS,EA}. Therefore, in this paper, we also
assume that the aether is regular across the sound horizon. Applying such conditions to the above solutions, we find that we   must set,
\begin{equation}
\label{CDb}
\tilde\chi_{a} =  \tilde\chi_{c}  = 0.
\end{equation}
It is interesting to
note that the $\hat C_2$ mode appearing in Eq.(\ref{EXP7}),  originally  from the zeroth-order perturbations $\bar{\chi}$, makes $\tilde\chi$ also singular across
the sound horizon for the case $l = 1$.  Therefore, in contrast to the case $c_{\phi} = \infty$, in the large $c_{\phi}$ case, this mode should be absent, once high-order
perturbations are considered. This can be understood as follows:   When   $c_{\phi} = \infty$ the sound horizon coincides with the universal horizon, 
and the regularity condition   of the aether across the sound horizon is no longer necessary. Instead, it is sufficient to require the singular behavior of the solutions  not be  worse than
$\left|\xi - \xi_{UH}\right|^{-1}$, as required  by Eq.(\ref{II6.b}). However, in the large but finite $c_{\phi}$ case, the location of the sound horizon is different from that of
the universal horizon, and  the regularity condition of the aether across the sound horizon must be imposed independently. As a result, in addition to the condition
(\ref{CDb}), we must also impose the condition,
\begin{equation}
\label{FFc}
C_2  = 0, \; (c_{\phi} \not= \infty).
\end{equation}

 Across the universal horizon, located at $\xi = \xi_{UH}$, where $\xi_{UH}$ is given by Eq.(\ref{new11}), we find that, to the leading order,  $\tilde\chi$ takes a similar form,
 as that given by Eq.(\ref{EXP7}). This is understandable, if we compare Eq.(\ref{new7}) with Eq.(\ref{new11}) and consider the fact that they coincide to the zeroth-order of
 $\epsilon$. Then, the asymptotic  condition (\ref{II6.b}) requires
 \begin{equation}
\label{CDc}
\tilde\chi'_{a} =  \tilde\chi'_{c}  = 0,
\end{equation}
for any given $l \ge 1$. Eqs.(\ref{CDa}), (\ref{CDb}) and (\ref{CDc}) are the conditions that the solutions $\tilde{\chi}$ must satisfy. However, these represent seven independent
conditions, imposed on $\tilde{\chi}$, which generically has only four integration constants, as can be seen from Eq.(\ref{new13}). Then, the system is overdetermined, and
the perturbation $\chi$ must be zero generically for any cases with a finite $c_{\phi}$, no matter how large it will be, as long as it is finite, $c_{\phi}  \not= \infty$.

\section{Conclusions}

In this paper, we have revisited the  problem of the existence of long static hairs  of black holes in theories without Lorentz symmetry in the case where the speed $c_{\phi}$ of the 
khronon field becomes infinitely large, $c_{\phi} = \infty$ (which is equivalent to $c_1+c_4=0$), so that   the sound horizon of the khronon field coincides with the universal horizon, and the boundary conditions at the sound 
horizon reduce to those given at the universal horizons. As a result, less boundary conditions are present in this 
extreme case in comparison with the case $c_{\phi} = $ finite. Then, it would be expected that static hairs might exist. However, 
we have shown analytically that even in this case static hairs still cannot exist. We also consider the cases in which $c_{\phi}$ is 
finite  but with  $c_{\phi} \gg 1$, and obtain the same conclusion.  

We would like to note that in the RN background our proof has been restricted to the case $c_3 = 0$.
Although it is not clear mathematically how to generalize the proof to the case with general values of $c_3$, it is not difficult to see that most likely  static hairs do not exist even for any given $c_3$.

\section*{\bf Acknowledgements}

This work was done partly when A.W. was visiting the State University of Rio de Janeiro (UERJ) and Yukawa Institute for Theoretical Physics (YITP), Kyoto University. 
He would like to thank both UERJ and YITP for their hospitality.
The work of A.W. is supported in part by  Ci\^encia Sem Fronteiras, No. A045/2013 CAPES, Brazil; National Natural Science Foundation of
China (NNSFC), Grant Nos. 11375153 and 11675145. The work of K.L. is supported  in part by   FAPESP No. 2012/08934-0, Brazil,  CAPES and CNPq, Brazil,
and NNSFC Nos.11573022 and 11375279. The  work of S.M.   is supported by Japan Society for the Promotion of Science (JSPS) Grants-in-Aid for Scientific
Research (KAKENHI) No. 24540256, No. 17H02890 and by World Premier International Research Center Initiative (WPI), MEXT, Japan. The work of T.Z is supported
in part by NNSFC, Nos.: 11675143, 11105120 and 11205133.

\section*{\bf Appendix A:  Another Derivation  of Solutions of Eq.(\ref{II8})}
\renewcommand{\theequation}{A.\arabic{equation}} \setcounter{equation}{0}

 In order to apply {\em the uniform asymptotic approximation method}  to construct analytical solution of $Y(\xi)$, let us first write the equation in the 
 standard form \cite{olver_asymptotic_1975, zhu_inflationary_2014, zhu_high-order_2016}
\begin{eqnarray}\label{eom_uniform}
Y'' = \left\{\lambda^2 \hat g(\xi) + q(\xi)\right\} Y,
\end{eqnarray}
where
\begin{eqnarray}\label{gplusq}
\lambda^2 \hat g(\xi) + q(\xi) = \frac{A(\xi)}{B(\xi)}.
\end{eqnarray}
Note that the above equation cannot determine the two functions $\lambda^2 \hat g(\xi)$ and $q(\xi)$ uniquely. A fundamental reason to introduce two of them is to have one extra degree of freedom, so we are allowed
to choose them in such a way that the error control functions, associated with the uniform asymptotic
approximation, can be  minimized. The convergence of these error control functions is very sensitive to their behaviors near the poles (singularities) of the functions $\lambda^2 \hat g(\xi)$ and $q(\xi)$, as shown in \cite{olver_asymptotic_1975, zhu_inflationary_2014, zhu_high-order_2016}. Let us first study the poles of $\lambda^2 \hat g(\xi)+q(\xi)$, which has two asymptotic limits
\begin{eqnarray}
\lambda^2 \hat g(\xi) + q(\xi) =
\begin{cases}
\frac{L^2}{\xi^2}, & \text{when} \;\; \xi \to 0^+\\
\frac{2L^2+3}{ 4\left(\xi_{UH}-\xi\right)^2}, & \text{when}\;\;  \xi \to \xi_{UH},
\end{cases}
\end{eqnarray}
where in the current case we have $\xi_{UH} = 4/3$.
Obviously the functions $\lambda^2 \hat g(\xi)$ and $q(\xi)$ may has two poles that located at $0^+$ and $4/3$, respectively. Both poles has order of $2$. According to the analysis given 
 in refs. \cite{olver_asymptotic_1975, zhu_inflationary_2014, zhu_high-order_2016}, in order to make the error control function finite near the poles, one has to choose
\begin{eqnarray}
q(\xi) =
\begin{cases}
- \frac{1}{4 \xi^2}, & \;\;\;\text{when}\;\; \xi \to 0^+,\\
- \frac{1}{4(\xi_{UH} -\xi)^2}, &\;\;\; \text{when}\;\; \xi \to \xi_{UH}.
\end{cases}
\end{eqnarray}
However,  in order to study the solution near both poles in a unified way, we can choose
\begin{eqnarray}
\label{A.1}
q(\xi) = - \frac{1}{4\xi^2}- \frac{1}{4(\xi_{UH}-\xi)^2},
\end{eqnarray}
thus
\begin{eqnarray}
\lambda^2 \hat g(\xi) = \frac{A(\xi)}{B(\xi)}  + \frac{1}{4\xi^2}+\frac{1}{4(\xi_{UH}-\xi)^2}.
\end{eqnarray}
The behaviors of the approximate solutions constructed in the uniform asymptotic approximation is also very sensitive to the turning points (zeros of function $\lambda^2 \hat g(\xi)$)
of Eq.(\ref{eom_uniform}). However, it is easy to show that  in the region $\xi \in (0,4/3)$, $\lambda^2 \hat g(\xi)$ does not vanish. Therefore, in the current case for the choice of
Eq.(\ref{A.1}) there are no turning points. Then, $\lambda^2 \hat g(\xi)$ is a totally regular function and thus the solution of Eq.(\ref{eom_uniform}) can be constructed by using the
Liouville-Green solutions,
\begin{eqnarray}
Y(\xi) &=& \frac{\alpha_1/\sqrt{2}}{(\hat g(\xi))^{1/4}} \exp{\left(- \int \sqrt{\hat g(\xi)} d\xi\right)}\nonumber\\
&& +\frac{\beta_1/\sqrt{2}}{(\hat g(\xi))^{1/4}}  \exp{\left( \int \sqrt{\hat g(\xi)} d\xi\right)}.
\end{eqnarray}
Note that in writing the above expression we simply set $\lambda = 1$ without loss of the generality.

Now let turn to  get the general solution of $\chi$ by solving Eq.(\ref{II10}), which is a non-homogeneous second-order ordinary differential equation. To find its general solutions, let us   first consider the solutions of the homogeneous part,
which reads,
\begin{eqnarray}
\chi''-\frac{3(1-4 r_0^4 \xi^3)}{2(1-\xi+r_0^4\xi^4)}\chi'-\frac{L^2}{\xi^2(1-\xi+r_0^4 \xi^4)}\chi=0.
\end{eqnarray}
Defining
\begin{eqnarray}
\chi(\xi) = (1-\xi+r_0^4 \xi^4)^{-3/4} u(\xi),
\end{eqnarray}
we find
\begin{eqnarray}
&& u''(\xi) +\frac{D(\xi)}{16\xi^2(1-\xi+r_o^4\xi^4)^2} u(\xi)=0,
\end{eqnarray}
where $D(\xi) \equiv 3\xi^2(1-8r_o^4\xi^2(6-5\xi+4r_o^4\xi^4))-16L^2(1-\xi+r_o^4\xi^4)$.
It is easy to show that this equation is exactly the same as that  for $Y(\xi)$.
Thus, the  solution of $u(\xi)$ has two independent branches in the region $\xi\in (0,\; \xi_{UH})$,   given, respectively,  by
\begin{eqnarray}
u_1(\xi) &=& \frac{1}{\sqrt{2}(\hat g(\xi))^{1/4}} \exp{\left(- \int \sqrt{\hat g(\xi)} d\xi\right)},\nonumber\\
 u_2(\xi) &=& \frac{1}{\sqrt{2}(\hat g(\xi))^{1/4}}  \exp{\left( \int \sqrt{\hat g(\xi)} d\xi\right)}.
\end{eqnarray}

To solve the inhomogeneous equation (\ref{II9}), we  assume that the solution  takes the form
\begin{eqnarray}
\chi_0(\xi) = c_1(\xi) \chi_1(\xi)+c_2(\xi) \chi_2(\xi),
\end{eqnarray}
where
\begin{eqnarray}
c_1(\xi) &=& - \int \frac{\chi_2(\xi)}{W(\xi)} \frac{H(\xi)}{2\xi^2 (1-\xi+r_0^4 \xi^4)^2}d\xi,\\
c_2(\xi) &=&  \int \frac{\chi_1(\xi)}{W(\xi)} \frac{H(\xi)}{2\xi^2 (1-\xi+r_0^4 \xi^4)^2}d\xi.
\end{eqnarray}
Here $W(\xi)$ is the Wronskian of $\chi_1$ and $\chi_2$ and is defined by
\begin{eqnarray}
W(\xi) = \chi_1 (\xi)\chi'_2(\xi)-\chi'_1(\xi) \chi_2(\xi).
\end{eqnarray}
Then the general solution of $\chi$ can be expressed as
\begin{eqnarray}
\chi(\xi) =  \chi_0(\xi) + \alpha_2 \chi_1(\xi) + \beta_2 \chi_2(\xi).
\end{eqnarray}

\subsection{Asymptotic behavior at spatial infinity}

At the spatial infinity, $\xi = r^{-1} = 0$, it is easy to see that
\begin{eqnarray}
\sqrt{\hat g(\xi)} \to \frac{\sqrt{L^2+1/4}}{\xi}.
\end{eqnarray}
Then,  we find
\begin{eqnarray}
&&H(\xi) \to \frac{\alpha_1}{\sqrt{2l+1}} \xi^{-l -2} + \frac{\beta_1}{\sqrt{2l+1}} \xi^{l-1},\nonumber\\
&&\chi_1(\xi) \to \frac{1}{\sqrt{2l+1}} \xi^{-l},\;\; \chi_2(\xi) \to \frac{1}{\sqrt{2l+1}} \xi^{l+1}.\nonumber\\
\end{eqnarray}
We also have $W=1$. Thus we obtain
\begin{eqnarray}
c_1(\xi) &\to& - \int  \frac{\xi^{l+1}}{\sqrt{2l+1}} \frac{1}{2}\left(\frac{\alpha_1}{\sqrt{2l+1}} \xi^{-l -4} \right.\nonumber\\
&& ~~~~~~~~ \left. + \frac{\beta_1}{\sqrt{2l+1}} \xi^{l-3}\right)d\xi \nonumber\\
&=&  \frac{\alpha_1}{4(2l+1)}\xi^{-2} - \frac{\beta_1}{2(4l^2-1)} \xi^{2l-1},\nonumber\\
c_2(\xi) &\to&  \int  \frac{\xi^{-l}}{\sqrt{2l+1}} \frac{1}{2}\left(\frac{\alpha_1}{\sqrt{2l+1}} \xi^{-l -4}\right.\nonumber\\
&& \left. ~~~~~~~ + \frac{\beta_1}{\sqrt{2l+1}} \xi^{l-3}\right)d\xi \nonumber\\
&=& - \frac{\alpha_1}{2(2l+1)(2l+3)}\xi^{-2l-3} - \frac{\beta_1}{4(2l+1)} \xi^{-2}.\nonumber\\
\end{eqnarray}
Then, we have,
\begin{eqnarray}
\chi_0(\xi) \to \frac{1}{\sqrt{2l+1}}\left[\frac{\alpha_1}{8l+12} \xi^{-l-2} - \frac{\beta_1}{8l-4} \xi^{l-1}\right]. ~~~~~~~~
\end{eqnarray}
Therefore, finally we find
\begin{eqnarray}
&& \chi(\xi) \to\frac{1}{\sqrt{2l+1}}\Bigg(\frac{\alpha_1}{8l+12} \xi^{-l-2} - \frac{\beta_1}{8l-4} \xi^{l-1}\nonumber\\
&& ~~~~~~~~~~~~~~~~ +\alpha_2  \xi^{-l}+ \beta_2 \xi^{l+1}\Bigg).
\end{eqnarray}
Then, the vanishing conditions of $\chi$   at $\xi\to 0^+$ requires,
\begin{eqnarray}
\label{alpha1_alpha_2}
&& (i) \; \alpha_1= \alpha_2=0,\; (l \ge 2), \nonumber\\
&& (ii) \;  \alpha_1= \alpha_2= \beta_1=0, \;  (l = 1).
\end{eqnarray}

\subsection{Asymptotic behavior at  the universal horizon}

At the universal horizon, we find
\begin{eqnarray}
\sqrt{\hat g(\xi)} \to \frac{\sqrt{(L^2+2)/2}}{ 4/3-\xi},
\end{eqnarray}
so that
\begin{eqnarray}
H(\xi) &\to& \frac{3\sqrt{3}}{8\sqrt{2}} \frac{1}{(L^2+2)^{1/4}} \Bigg[\alpha_1 3^{\sqrt{1+L^2/2}}\nonumber\\
&& \times \left(\xi_{UH}-\xi\right)^{\sqrt{1+L^2/2}}\nonumber\\
&& +\beta_1 3^{-\sqrt{1+L^2/2}} \left(\xi_{UH}-\xi\right)^{-\sqrt{1+L^2/2}}\Bigg],\nonumber\\
\chi_1(\xi) &\to & \frac{4\times 2^{1/4}}{3\sqrt{6}} \frac{3^{\sqrt{1+L^2/2}}}{(1+L^2/2)^{1/4}} \left(\xi_{UH}-\xi\right)^{\sqrt{1+L^2/2}-1},\nonumber\\
\chi_2(\xi) &\to&  \frac{4\times 2^{1/4}}{3\sqrt{6}} \frac{3^{-\sqrt{1+L^2/2}}}{(1+L^2/2)^{1/4}} \left(\xi_{UH}-\xi\right)^{-\sqrt{1+L^2/2}-1}.\nonumber\\
\end{eqnarray}
Here, we have
\begin{eqnarray}
W(\xi)&=&\chi_(\xi) \chi'_2(\xi)-\chi'_1(\xi) \chi_2(\xi)\nonumber\\
&=& \frac{64}{27(2+L^2)^{3/2}} \left(\xi_{UH}-\xi\right)^{-3}.
\end{eqnarray}
Then, we find
\begin{eqnarray}
&& c_1(\xi) \to -  \frac{3}{32\times 2^{3/4}} \Bigg[\alpha_1\left(\xi_{UH}-\xi\right)^{-1}\nonumber\\
&& ~~~~~~~~~~ +\beta_1 \frac{3^{-2\sqrt{1+L^2/2}}}{2\sqrt{1+L^2/2}} \left(\xi_{UH}-\xi\right)^{-2\sqrt{1+L^2/2}-1}\Bigg],\nonumber\\
&& c_2 (\xi) \to  \frac{3}{32\times 2^{3/4}} \Bigg[\beta_1\left(\xi_{UH}-\xi\right)^{-1}\nonumber\\
&& ~~~~~~~~~~ +\alpha_1 \frac{3^{2\sqrt{1+L^2/2}}}{2\sqrt{1+L^2/2}} \left(\xi_{UH}-\xi\right)^{2\sqrt{1+L^2/2}-1}\Bigg], \nonumber\\
\end{eqnarray}
and
\begin{eqnarray}
&& \chi_0(\xi) \to - \frac{L^2+\sqrt{1+L^2/2}+2}{4\sqrt{3}(2L^4+7L^2+6)}\nonumber\\
&&  ~~~~~~~~~~~ \times  \Bigg[\alpha_1 3^{\sqrt{1+L^2/2}} \left(\xi_{UH}-\xi\right)^{\sqrt{1+L^2/2}-2}\nonumber\\
&& ~~~~~~~~~~~~~ - \beta_1 3^{-\sqrt{1+L^2/2}} \left(\xi_{UH}-\xi\right)^{-\sqrt{1+L^2/2}-2} \Bigg].\nonumber\\
\end{eqnarray}
Finally, we find that  the solution of $\chi(\xi)$ near the universal horizon takes the form, 
\begin{eqnarray}
\chi(\xi) &=&  \chi_(\xi)+\alpha_2 \chi_1(\xi) + \beta_2 \chi_2 (\xi)\nonumber\\
&\to & - \frac{L^2+\sqrt{1+L^2/2}+2}{4\sqrt{3}(2L^4+7L^2+6)}\nonumber\\
&& \times  \Bigg[\alpha_1 3^{\sqrt{1+L^2/2}} \left(\xi_{UH}-\xi\right)^{\sqrt{1+L^2/2}-2}\nonumber\\
&&  - \beta_1 3^{-\sqrt{1+L^2/2}} \left(\xi_{UH}-\xi\right)^{-\sqrt{1+L^2/2}-2} \Bigg]\nonumber\\
&&+\alpha_2  \frac{4\times 2^{1/4}}{3\sqrt{6}} \frac{3^{\sqrt{1+L^2/2}}}{(1+L^2/2)^{1/4}} \nonumber\\
&& \times \left(\xi_{UH}-\xi\right)^{\sqrt{1+L^2/2}-1}\nonumber\\
&& +\beta_2 \frac{4\times 2^{1/4}}{3\sqrt{6}} \frac{3^{-\sqrt{1+L^2/2}}}{(1+L^2/2)^{1/4}}\nonumber\\
&& \times  \left(\xi_{UH}-\xi\right)^{-\sqrt{1+L^2/2}-1}.\nonumber\\
\end{eqnarray}
Then, the requirement that $\chi(\xi)$ be finite as $\xi \to \xi_{UH}$ leads to
\begin{eqnarray}
\beta_1 =0, \;\;\; \beta_2=0.
\end{eqnarray}
Together with Eq.(\ref{alpha1_alpha_2}), we finally obtain
\begin{eqnarray}
\alpha_1=0, \;\; \alpha_2 =0 ,\;\; \beta_1 =0,\;\; \beta_2 =0,
\end{eqnarray}
that is, the only solution of $\chi$ that satisfies the boundary conditions is the trivial one,
\begin{equation}
\chi(\xi) = 0.
\end{equation}

\section*{\bf Appendix B: Formulas for Large $c_{\phi}$ Expansions}
\renewcommand{\theequation}{B.\arabic{equation}} \setcounter{equation}{0}

The coefficients of Eq.(\ref{new13}) are given by
 \begin{eqnarray}
 \label{new14}
\hat A(\xi)&=&\frac{1}{c_{\phi}^2} \xi^4U^4\left(V^2-c_\phi^2U^2\right),\nonumber\\
\hat B_l(\xi)&=&  \frac{1}{c_{\phi}^2}\Big\{-2\xi^4U^3V^2U''-\xi^4U^4VV''-2\xi^4U^2V^2U'^2\nonumber\\
&&-2\xi^2U^4V^2-4\xi^4U^3U'VV'-4\xi^3U^4VV'\nonumber\\
&&-8\xi^3U^3U'V^2+l^2\xi^2U^2V^2-\xi^4U^5U''\nonumber\\
&&-2\xi^4U^3V^2U''-c_\phi^2\left(-6\xi^4U^4U'^2-12\xi^3U^5U'\right.\nonumber\\
&&-3\xi^4U^5U''+2l^2\xi^2U^4+6\xi^2U^4V^2\nonumber\\
&&\left.-3\xi^4U^4VV''\right)\Big\},\nonumber\\
\hat C_l(\xi)&=&-\frac{l^2}{c_{\phi}^2}\Big\{2\xi UU'V^2+\xi^2UV^2U''+2\xi^2UVU'V'\nonumber\\
&&+\xi^2U^2U'^2+\xi^2U^3U''-c_\phi^2\left(-l^2U^2+\xi^2U^3U''\right.\nonumber\\
&&+3\xi^2U^2U'^2+10\xi U^3U'+2U^4-2U^2V^2\nonumber\\
&&\left.+\xi^3U^2VV''\right)\Big\}.
 \end{eqnarray}
Expanding the above expressions  in terms of $\epsilon$,  we find
 \begin{eqnarray}
 \label{new14a}
\hat
A(\xi)&=&\frac{\xi^4(3\xi-4)^4(3\xi^2+8\xi+16)^2}{16777216}\Bigg[256 (\xi -1)
-27\xi ^4 \nonumber\\
&&+9 \epsilon^2 \xi ^2 \left(3 \xi ^2+32 \sqrt{3} V_2\right)\Bigg]
+ {\cal{O}}\left(\epsilon^4\right),\nonumber\\
\hat B_l(\xi)&=&-\frac{\xi^2(3\xi-4)^2}{8388608}\Big\{\left[\xi
(3 \xi +8)+16\right] (4-3 \xi )^2\nonumber\\
&&\times \big[256 l^2 (3 \xi^2 +8\xi+16)\nonumber\\
&& -3 \xi (3 \xi (81 \xi ^2 (\xi
(3 \xi +8)+16)-512)-4096)\big]\nonumber\\
&& +\epsilon^2 \xi ^2 \Big[\xi (3 \xi(-1152 l^2 (\xi  (3 \xi +8)+16)\nonumber\\
&& +9 \xi  (\xi  (3 \xi (279 \xi  (\xi  (3 \xi+8)+16)\nonumber\\
&& -3584)-18688)-30720)+323584)\nonumber\\
&&-65536)+48 \sqrt{3}(3 \xi  (\xi  (3 \xi +8)+16) (2(81 \xi^4\nonumber\\
&&-576 \xi+512) V_2'+\xi(27 \xi ^4-256 \xi \nonumber\\
&&+256) V_2'')-8(128 l^2 (\xi  (3 \xi +8)+16)\nonumber\\
&&-3(\xi\left(135 \xi ^3-224\right) (\xi  (3 \xi
+8)+16)\nonumber\\
&&+2048)) V_2)-131072\Big]\Big\} + {\cal{O}}\left(\epsilon^4\right),\nonumber\\
\hat
C_l(\xi)&=&-\frac{l}{16384(3\xi^2+8\xi+16)}\Big\{(3\xi^2+8\xi\nonumber\\
&&+16) (4-3 \xi )^2 [64 l^2 (3 \xi^2 +8\xi+16)\nonumber\\
&&+3 \xi(\xi  (896-3 \xi  (27 \xi  (\xi  (3 \xi+8)+16)-64))\nonumber\\
&&+2048)-2048]\nonumber\\
&&+3\epsilon^2 \xi ^2 \Big[8 \sqrt{3} (\xi  (3 \xi +8)+16)
\big(\xi((297 \xi ^4-2048 \xi \nonumber\\
&&+1792) V_2'+\xi \left(27\xi ^4-256 \xi +256\right) V_2'')\nonumber\\
&&-16 \left(16 l^2-81 \xi^4+336 \xi -224\right) V_2\big)\nonumber\\
&&+9 (3 \xi -4) (\xi  (\xi  (45 \xi(3\xi(\xi
+4)+32)+1216)\nonumber\\
&&+384)-1536) \xi ^2\Big]\Big\} + {\cal{O}}\left(\epsilon^4\right).
 \end{eqnarray}

The coefficients $\hat P_i$ and $\hat Q_i$ appearing in Eq.(\ref{EXP2}) are given by
 \begin{eqnarray}
 \label{EXP3}
 \hat P_3&=&\frac{8}{\xi}+\frac{36}{3\xi-4}+\frac{48+36\xi}{3\xi^2+8\xi+16},\nonumber\\
 \hat P_2&=&\frac{-512l^2}{\xi^2(27\xi^4-256\xi+256)}+\frac{6}{\xi^2(3\xi^2+8\xi+16)^2}\nonumber\\
 &&\times(3\xi-4)^{-2}(8192-24576\xi-8192\xi^2+1536\xi^3\nonumber\\
 &&+21168\xi^4+10584\xi^5+3969\xi^6),\nonumber\\
 \hat P_1&=&-\frac{1024l^2(45\xi^3+60\xi^2+80\xi-64)}{\xi^3(3\xi-4)^3(3\xi^2+8\xi+16)^2}\nonumber\\
 &&+\frac{12(3\xi^2+8\xi+16)^{-3}}{\xi^2(3\xi-4)^3}(393216-229376\xi\nonumber\\
 &&-442368\xi^2-1022976\xi^3-6912\xi^4+404352\xi^5\nonumber\\
 &&+489888\xi^6+183708\xi^7+45927\xi^8),
 \end{eqnarray}
 \begin{eqnarray}
 \hat P_0&=&\frac{65536l^4}{\xi^4(4-3\xi)^4(3\xi^2+8\xi+16)^2}\nonumber\\
 &&+\frac{1024l^2(4-3\xi)^{-4}}{\xi^4(3\xi^2+8\xi+16)^4}(2048-6144\xi-2688\xi^2\nonumber\\
 &&-576\xi^3+3888\xi^4+1944\xi^5+729\xi^6),\\
 \hat Q_4&=&\frac{-27\xi^4-288\sqrt{3}\xi^2V_2}{27\xi^4-256\xi+256},\nonumber\\
 \hat Q_3&=&-\frac{576\sqrt{3}\xi^2V'_2}{27\xi^4-256\xi+256}-\frac{\xi(864\xi^2+4608\sqrt{3}V_2)}{(3\xi-4)^3}\nonumber\\
 &&\times\frac{63\xi^3+84\xi^2+112\xi-192}{3\xi^2+8\xi+16}, \nonumber\\
 \hat Q_2&=&-\frac{576 \sqrt{3} \xi ^2 V_2''}{27 \xi ^4-256 \xi +256}-\frac{576
   \sqrt{3}(3 \xi -4)^{-3} \xi  V_2'}{ \left(3 \xi ^2+8 \xi +16\right)^2}\nonumber\\
 &&\times(153 \xi ^3+204 \xi ^2+272 \xi -512)\nonumber\\
 &&+\frac{2(3\xi-4)^{-4}}{(3\xi^2+8\xi+16)^3}[3456 \left(3 l^2+650\right) \xi ^4\nonumber\\
 &&+27648 \left(l^2+150\right) \xi ^3+6144
   \left(9 l^2-662\right) \xi ^2\nonumber\\
 &&+96 \sqrt{3} (512 l^2 (3 \xi
   ^2+8 \xi +16)-3 (8991 \xi ^6\nonumber\\
 &&+23976 \xi ^5+47952 \xi ^4-17280
   \xi ^3-46848 \xi ^2\nonumber\\
 &&-94208 \xi +69632)) V_2(\xi )-330237 \xi^8\nonumber\\
 &&-880632 \xi ^7-1761264 \xi ^6+1036800 \xi ^5\nonumber\\
 &&+65536 \xi +131072], \nonumber\\
 \hat Q_1&=&-\frac{288 \sqrt{3} \xi ^2 V_2{}^{(3)}}{27 \xi ^4-256 \xi +256}-\frac{5760 \sqrt{3}  \xi(3\xi -4)^{-3}  V_2''}{\left(3 \xi ^2+8 \xi +16\right)^2}\nonumber\\
 &&\times\left(9 \xi ^3+12 \xi ^2+16 \xi -32\right)-\frac{4(3\xi-4)^{-5}\xi^{-1}}{(3\xi^2+8\xi+16)^3}\nonumber\\
 &&\times\{-155520 l^2 \xi ^5-207360 l^2 \xi ^4-276480 l^2 \xi ^3\nonumber\\
 &&+663552 l^2 \xi^2+48 \sqrt{3} (3 \xi -4) \xi  [3 (4779 \xi^6\nonumber\\
 &&+12744 \xi^5+25488 \xi ^4-13056 \xi ^3-30208 \xi ^2\nonumber\\
 &&-57344 \xi +49152)-512
   l^2 \left(3 \xi ^2+8 \xi +16\right)] V_2'\nonumber\\
 &&-768 \sqrt{3}[256 l^2 \left(9 \xi ^3+12 \xi ^2+16 \xi -32\right)\nonumber\\
 &&-3 (3645
   \xi ^7+4860 \xi ^6+6480 \xi ^5-21816 \xi ^4\nonumber\\
 &&-6048 \xi ^3-8064 \xi
   ^2+27136 \xi -8192)] V_2\nonumber\\
 &&+1423737 \xi ^9+1898316 \xi
   ^8+2531088 \xi ^7\nonumber\\
 &&-12021696 \xi ^6-2343168 \xi ^5-3124224 \xi^4\nonumber\\
 &&+23261184 \xi ^3-12337152 \xi ^2-524288 \xi \nonumber\\
 &&+1048576\},\nonumber\\
 \hat Q_0&=&\frac{27648 l^2(3 \xi -4)^{-5}}{ \left(3 \xi ^2+8 \xi +16\right)^4}(135 \xi ^5+540 \xi ^4+1440 \xi ^3\nonumber\\
 &&+1216 \xi ^2+384 \xi -1536)\nonumber\\
 &&+\frac{393216\sqrt{3} l^2(3 \xi -4)^{-6}V_2}{\xi^2 \left(3 \xi ^2+8 \xi +16\right)^3}(81 \xi ^4-336 \xi\nonumber\\
 && +224-16 l^2)+\frac{24576\sqrt{3} l^2(3 \xi -4)^{-5}V'_2}{\xi \left(3 \xi ^2+8 \xi +16\right)^3}(99 \xi ^3\nonumber\\
 &&+132 \xi ^2+176 \xi -448)\nonumber\\
 &&+\frac{24576 \sqrt{3} l^2 V_2''}{(4-3 \xi )^4 \left(3 \xi ^2+8 \xi +16\right)^2}.
 \end{eqnarray}


\end{document}